\pdfoutput=1

\documentclass[aps,prd,notitlepage,twoside,twocolumn,preprintnumbers,showpacs,showkeys,byrevtex,floatfix,nofootinbib,amsmath,amssymb]{revtex4-1}
\usepackage{amscd,array,dsfont,graphicx,kbordermatrix,mathtools,multirow,tensor}
\usepackage[bookmarksnumbered,bookmarksopen=true,bookmarksopenlevel=1,linktocpage,pdfstartview=FitH]{hyperref}
\usepackage[all]{hypcap}

\allowdisplaybreaks[1]

\newcolumntype{M}[1]{>{$}{#1}<{$}}

\renewcommand{\kbldelim}{(}
\renewcommand{\kbrdelim}{)}

\DeclarePairedDelimiter{\ket}{\lvert}{\rangle}
\DeclarePairedDelimiter{\bra}{\langle}{\rvert}

\DeclareMathOperator{\Ber}{Ber}
\DeclareMathOperator{\tr}{tr}
\DeclareMathOperator{\str}{str}
\DeclareMathOperator{\Det}{Det}
\DeclareMathOperator{\sdet}{sdet}
\DeclareMathOperator{\sDet}{sDet}

\newcommand{\bp}{\bullet}
\newcommand{\eps}{\varepsilon}

\newcommand{\cmplxs}{\mathds{C}}

\newcommand{\half}{\tfrac{1}{2}}
\newcommand{\rep}[1]{\ensuremath{\mathbf{#1}}}
\newcommand{\braket}[2]{\langle#1|#2\rangle}
\newcommand{\ketbra}[2]{\lvert#1\rangle\!\langle#2\rvert}
\newcommand{\llbracket}{[\![}
\newcommand{\rrbracket}{]\!]}

\begin{document}

\title{Superqubits}
\author{L. Borsten}
\email[]{leron.borsten@imperial.ac.uk}
\author{D. Dahanayake}
\email[]{duminda.dahanayake@imperial.ac.uk}
\author{M. J. Duff}
\email[]{m.duff@imperial.ac.uk}
\author{W. Rubens}
\email[]{william.rubens06@imperial.ac.uk}
\affiliation{Theoretical Physics, Blackett Laboratory, Imperial College London, London SW7 2AZ, United Kingdom}
\date{\today}

\begin{abstract}

We provide a supersymmetric generalization of $n$ quantum bits  by extending the local operations and classical communication entanglement equivalence group $[SU(2)]^n$ to the supergroup $[uOSp(1|2)]^n$ and the stochastic local operations and classical communication equivalence group $[SL(2,\mathds{C})]^n$ to the supergroup $[OSp(1|2)]^n$. We introduce the appropriate supersymmetric generalizations of the conventional entanglement measures for the cases of $n=2$ and $n=3$. In particular, super-Greenberger-Horne-Zeilinger states are characterized by a nonvanishing superhyperdeterminant.

\end{abstract}

\pacs{11.30.Pb, 03.65.Ud, 03.67.Mn}

\keywords{qubit, entanglement, hyperdeterminant}

\preprint{Imperial/TP/2009/mjd/2}

\maketitle

\section{Introduction}\label{sec:intro}

The question of computable entanglement measures for arbitrary quantum systems is, to a large extent, an open one. However, substantial progress has been made utilizing the paradigms of  \emph{local operations and classical communication} (LOCC) and \emph{stochastic local operations and classical communication} (SLOCC). For example,  2-qubit and 3-qubit systems both admit concise, but nontrivial, SLOCC classifications which reveal a number of important qualitative features of multipartite entanglement \cite{Coffman:1999jd, Dur:2000, Acin:2001, Miyake:2002, Miyake:2003, Borsten:2009yb}. In particular, 2-qubit Bell states and 3-qubit Greenberger-Horne-Zeilinger (GHZ) states are characterized, respectively, by  nonvanishing determinant and hyperdeterminant.

Here we propose a supersymmetric generalization of the qubit,  the  \emph{superqubit}. We proceed by extending the $n$-qubit SLOCC equivalence group $[SL(2,\mathds{C})]^n$ and the LOCC equivalence group $[SU(2)]^n$  to the supergroups $[OSp(1|2)]^n$ and $[uOSp(1|2)]^n$, respectively. A single superqubit forms a 3-dimensional representation of $OSp(1|2)$ consisting of two commuting  ``bosonic'' components and one anticommuting ``fermionic'' component. For $n=2$ and $n=3$ we introduce the appropriate supersymmetric generalizations of the conventional entanglement measures. In particular, super-Bell and super-GHZ states are characterized,  respectively, by  nonvanishing superdeterminant (distinct from the Berezinian) and superhyperdeterminant\footnote{The present work was in part inspired by the construction of the superhyperdeterminant in \cite{Castellani:2009pi}.}.

This mathematical construction seems a very natural one. Moreover, from a physical point of view, it makes contact with various condensed-matter systems. For example, the three-dimensional representation of $OSp(1|2)$ is encountered in the supersymmetric $t$-$J$ model where it describes spinons and holons on a one-dimensional lattice \cite{Wiegmann:1988, Sarkar:1991, PhysRevB.46.9234, Essler:1992nk, Mavromatos:1999xj}. It also shows up in the quantum Hall effect \cite{Hasebe:2004hy} and Affleck-Kennedy-Lieb-Tasaki models of superconductivity \cite{ Arovas:2009dx}.

In order to facilitate the introduction of a super Hilbert space, super LOCC and superqubits in \autoref{sec:superqubits}, we first recall some familiar properties of ordinary Hilbert space, LOCC, and qubits in \autoref{sec:qubits}. Similarly, in order to discuss the superentanglement of two and three superqubits in \autoref{sec:superentanglement}, we first review the ordinary entanglement of two and three qubits in \autoref{sec:entanglement}.

\section{Qubits}
\label{sec:qubits}

\subsection{Hilbert space}

  A complex Hilbert space $\mathcal{H}$ is equipped with a one-to-one map into its dual space $\mathcal{H}^\dag$,
\begin{equation}
\begin{split}
{}^\dag:\mathcal{H}&\to\mathcal{H}^\dag,\\
         \ket{\psi}&\mapsto (\ket{\psi})^\dag:=\bra{\psi}
\end{split}
\end{equation}
which defines an inner product $\braket{\psi}{\phi}$ and satisfies the following properties:
\begin{enumerate}
\item For all $\ket{\psi}, \ket{\phi}\in \mathcal{H}$, and any complex number $\alpha$ we have,
\begin{equation}
\begin{split}
(\alpha\ket{\psi})^\dag&=\bra{\psi}\alpha^*,\\
(\ket{\psi}+ \ket{\phi})^\dag&=\bra{\psi}+ \bra{\phi}.
\end{split}
\end{equation}
\item For all $\ket{\psi}, \ket{\phi}\in \mathcal{H}$,
\begin{equation}
\braket{\psi}{\phi}^*=\braket{\phi}{\psi}.
\end{equation}
\item For all $\ket{\psi}\in \mathcal{H}$,
\begin{equation}
\braket{\psi}{\psi}\geq 0
\end{equation}
with equality holding if and only if $\ket{\psi}$ is the null vector.
\end{enumerate}
	
In particular a qubit lives in the two-dimensional complex Hilbert space $\mathds{C}^2$. An arbitrary $n$-qubit system is then simply a vector in the $n$-fold tensor product Hilbert space $\mathds{C}^2\otimes\dotsb\otimes\mathds{C}^2=[\mathds{C}^2]^n$.

\subsection{LOCC and SLOCC}

Two states are said to be LOCC equivalent if and only if they  may be transformed into one another with certainty using LOCC protocols.  Reviews of the LOCC paradigm and entanglement measures may be found in \cite{Plenio:2007,Horodecki:2007}.  It is well known that two states of a composite system are LOCC equivalent if and only if they are related by the group of local unitaries (which we will refer to as the \emph{LOCC equivalence group}), unitary transformations that factorize into separate transformations on the component parts \cite{Bennett:1999}.  In the case of $n$ qubits  the group of local unitaries is given  (up to a global phase) by $[SU(2)]^n$.

Similarly, two quantum states are said to be SLOCC equivalent if and only if they may be transformed into one another with some \emph{nonvanishing probability} using LOCC operations \cite{Bennett:1999, Dur:2000}.  The set of SLOCC transformations relating equivalent states forms a group (which we will refer to as the \emph{SLOCC equivalence group}).  For $n$ qubits the SLOCC equivalence group is given  (up to a global complex factor) by the $n$-fold tensor product, $[SL(2, \mathds{C})]^n$, one factor for each qubit \cite{Dur:2000}. Note, the LOCC equivalence group forms a compact subgroup of the larger SLOCC equivalence group.

The Lie algebra $\mathfrak{sl}(2)$ may be conveniently summarized as
\begin{equation}\label{eq:slbrackets}
\left[ P_{A_1A_2}, P_{A_3A_4} \right]       = 2\eps_{(A_1(A_3} P_{A_4)A_2)}
\end{equation}
where $A=0,1$, and throughout this paper we use ``strength one'' (anti)symmetrization, so that
\begin{equation}
X_{(A_1A_2)}\equiv\half(X_{A_1A_2}+X_{A_2A_1}).
\end{equation}
We permit the  indices to be raised/lowered by the $SL(2, \mathds{C})$--invariant epsilon tensors according to the rules:
\begin{align}\label{eq:SLmetric}
V_{A_1}& = \eps_{A_1A_2} V^{A_2} & V^{A_1} &=  \eps^{A_1A_2} V_{A_2},
\end{align}
where we adopt the following conventions
\begin{equation}
\eps_{A_1A_2}= -\eps^{A_1A_2}, \quad \eps_{A_1A_2}\eps^{A_2A_3}=\delta_{A_1}^{A_3}.
\end{equation}
Consequently,
\begin{equation}
U^AV_A=-U_AV^A.
\end{equation}

The compact subalgebra $\mathfrak{su}(2)$ is given by
\begin{equation}
\mathfrak{su}(2):=\{X\in\mathfrak{sl}(2)|X^\dag=-X\}.
\end{equation}
An arbitrary element $X\in\mathfrak{su}(2)$ may be written as
\begin{equation}
X=\xi_i A_i ,
\end{equation}
where $\xi_i\in\mathds{R}$ and
\begin{equation}
\begin{gathered}
A_1=\tfrac{i}{2}(P_{00}- P_{11}),\\
A_2=\half(P_{00}+ P_{11}),\\
A_3=iP_{01},\\
A_{i}^{\dag}=-A_{i}.
\end{gathered}
\end{equation}

\subsection{One qubit}

The one-qubit system (Alice) is described by the state
\begin{equation}
\ket{\Psi} = a_{A}\ket{A},
\end{equation}
and the Hilbert space has dimension $2$. The SLOCC equivalence group is $SL(2, \mathds{C})_{A}$, under which $a_{A}$ transforms as a \textbf{2}.

The norm squared $\braket{\Psi}{\Psi}$ is given by
\begin{equation}
\langle \Psi | \Psi \rangle = \delta^{A_1A_2} a_{A_1}^{*}a_{A_2}
\end{equation}
and is invariant under $SU(2)_{A}$.
The one-qubit density matrix is given by
\begin{equation}
\begin{split}
\rho&:=\ketbra{\Psi}{\Psi}\\
&= a_{A_1}a_{A_2}^{*}\ketbra{{A_1}}{{A_2}}.
\end{split}
\end{equation}
The norm squared is then given by
\begin{equation}
\braket{\Psi}{\Psi}=\tr(\rho).
\end{equation}
Unnormalized pure state density matrices satisfy
\begin{equation}\label{eq:densitycondition}
\rho^2 = \tr(\rho)\rho.
\end{equation}

\subsection{Two qubits}

The two-qubit system (Alice and Bob) is described by the state
\begin{equation}
\ket{\Psi} = a_{AB}\ket{AB},
\end{equation}
and the Hilbert space has dimension $2^2=4$.   The SLOCC equivalence group is $SL(2, \mathds{C})_{A} \times SL(2, \mathds{C})_{B}$ under which $a_{AB}$ transforms as a $\mathbf{(2,2)}$.

The norm squared $\braket{\Psi}{\Psi}$ is given by
\begin{equation}
\langle \Psi | \Psi \rangle = \delta^{A_1A_2} \delta^{B_1B_2}a_{A_1B_1}^{*}a_{A_2B_2}.
\end{equation}
and is invariant under $SU(2)_{A} \times SU(2)_{B}$.
The two-qubit density matrix is given by
\begin{equation}
\begin{split}
\rho&:=\ketbra{\Psi}{\Psi}\\
&= a_{A_1B_1}a_{A_2B_2}^{*}\ketbra{{A_1B_1}}{{A_2B_2}}.
\end{split}
\end{equation}
The reduced density matrices are defined using the partial trace
\begin{equation}
\begin{split}
\rho_{A}=\tr_B \ket{\Psi}\bra{\Psi},\\
\rho_{B}=\tr_A \ket{\Psi}\bra{\Psi},\\
\end{split}
\end{equation}
or
\begin{equation}
\begin{split}
(\rho_{A})_{A_{1}A_{2}}&=\delta^{B_{1}B_{2}}a_{A_{1}B_{1}}a^*_{A_{2}B_{2}}, \\
(\rho_{B})_{B_{1}B_{2}}&=\delta^{A_{1}A_{2}}a_{A_{1}B_{1}}a^*_{A_{2}B_{2}}.
\end{split}
\end{equation}

\subsection{Three qubits}

The three-qubit system (Alice, Bob, Charlie) is described by the state
\begin{equation}
\ket{\Psi} = a_{ABC}\ket{ABC},\label{eq:hypermatrix}
\end{equation}
and the Hilbert space has dimension $2^3=8$.  The SLOCC equivalence group is $SL(2, \mathds{C})_{A} \times SL(2, \mathds{C})_{B} \times SL(2, \mathds{C})_{C}$ under which $a_{ABC}$ transforms as a $\mathbf{(2,2,2)}$.

The norm squared $\braket{\Psi}{\Psi}$ is given by
\begin{equation}
\langle \Psi | \Psi \rangle = \delta^{A_1A_2} \delta^{B_1B_2}\delta^{C_1C_2}a_{A_1B_1C_1}^{*}a_{A_2B_2C_2}
\end{equation}
and is invariant under $SU(2)_{A} \times SU(2)_{B} \times SU(2)_{C}$.
The three-qubit density matrix is given by
\begin{equation}
\begin{split}
\rho&:=\ketbra{\Psi}{\Psi}\\
&= a_{A_1B_1C_1}a_{A_2B_2C_2}^{*}\ketbra{{A_1B_1C_1}}{{A_2B_2C_2}}.
\end{split}
\end{equation}
The singly reduced density matrices are defined using the partial trace
\begin{equation}
\begin{split}
\rho_{AB}&=\tr_{C}\ket{\Psi}\bra{\Psi}, \\
\rho_{BC}&=\tr_{A}\ket{\Psi}\bra{\Psi}, \\
\rho_{CA}&=\tr_{B}\ket{\Psi}\bra{\Psi},
\end{split}
\end{equation}
or
\begin{equation}
\begin{split}
(\rho_{AB})_{A_1A_2B_1B_2}&=\delta^{C_1C_2}a_{A_1B_1C_1}a^{*}_{A_2B_2C_2}, \\
(\rho_{BC})_{B_1B_2C_1C_2}&=\delta^{A_1A_2}a_{A_1B_1C_1}a^{*}_{A_2B_2C_2}, \\
(\rho_{CA})_{C_1C_2A_1A_2}&=\delta^{B_1B_2}a_{A_1B_1C_1}a^{*}_{A_2B_2C_2}.
\end{split}
\end{equation}
The doubly reduced density matrices are defined using the partial traces
\begin{equation}
\begin{split}
\rho_A&=\tr_{BC}\ket{\Psi}\bra{\Psi}, \\
\rho_B&=\tr_{CA}\ket{\Psi}\bra{\Psi}, \\
\rho_C&=\tr_{AB}\ket{\Psi}\bra{\Psi},
\end{split}
\end{equation}
or
\begin{equation}
\begin{split}
(\rho_A)_{A_1A_2}&=\delta^{B_1B_2}\delta^{C_1C_2}a_{A_1B_1C_1}a^{*}_{A_2B_2C_2}, \\
(\rho_B)_{B_1B_2}&=\delta^{C_1C_2}\delta^{A_1A_2}a_{A_1B_1C_1}a^{*}_{A_2B_2C_2}, \\
(\rho_C)_{C_1C_2}&=\delta^{A_1A_2}\delta^{B_1B_2}a_{A_1B_1C_1}a^{*}_{A_2B_2C_2}.
\end{split}
\end{equation}

\section{Entanglement}
\label{sec:entanglement}

\subsection{Two qubits}

For two qubits there are only two distinct SLOCC entanglement classes - two qubits are either entangled or not.  The two classes are distinguished by the SLOCC invariant, $\det a_{AB}$. For separable states $\det a_{AB}=0$, while it is nonzero for any entangled state.

There are two independent $[SU(2)]^2$ invariants, the norm $\braket{\Psi}{\Psi}^{1/2}$ and the 2-tangle $\tau_{AB}$ \cite{Linden:1997qd, Coffman:1999jd},
\begin{equation}
\tau_{AB}=4 \det \rho_A = 4 \det \rho_B = 4 |\det a_{AB}|^2.
\label{eq:2-tangle}
\end{equation}\\
The 2-tangle is maximized, $ \tau_{AB}=1$, by the Bell state:
\begin{equation}
\ket{\Psi}_{\textrm{Bell}}=\tfrac{1}{\sqrt{2}}(\ket{00}+\ket{11}).
\end{equation}

\subsection{Three qubits}\label{sec:3qubitentanglement}

For three qubits there are six distinct SLOCC entanglement classes \cite{Dur:2000, Miyake:2002, Miyake:2003, Borsten:2009yb}. These classes and their  representative states are summarized as follows:
\begin{description}
  \item[Separable] Zero entanglement orbit for completely factorisable product states,
      \begin{equation}A\text{-}B\text{-}C:\quad\ket{000}.\end{equation}
  \item[Biseparable] Three classes of bipartite entanglement
      \begin{equation}
      \begin{split}
      A\text{-}BC:\quad\ket{010}+\ket{001},\\
      B\text{-}CA:\quad\ket{100}+\ket{001},\\
      C\text{-}AB:\quad\ket{010}+\ket{100}.
      \end{split}
      \end{equation}
  \item[W] Three-way entangled states that do not maximally violate Bell-type inequalities in the same way as the GHZ class discussed below. However, they are robust in the sense that tracing out a subsystem generically results in a bipartite mixed state that is maximally entangled under a number of criteria \cite{Dur:2000},
      \begin{equation}\text{W}:\quad\ket{100}+\ket{010}+\ket{001}.\end{equation}
  \item[GHZ] Genuinely tripartite entangled Greenberger-Horne-Zeilinger \cite{Greenberger:1989} states. These maximally violate Bell's inequalities but, in contrast to class W, are fragile under the tracing out of a subsystem since the resultant state is completely unentangled,
      \begin{equation}\text{GHZ}:\quad\ket{000}+\ket{111}.\end{equation}
\end{description}

The six classes may be distinguished either by appealing to simple arguments concerning the conservation of reduced density
matrix ranks  as in \cite{Dur:2000} or by considering the vanishing or not of five algebraically independent covariants/invariants as in \cite{Borsten:2009yb}. For our purposes it is more convenient to follow the latter approach as it better facilitates our supersymmetric extension. The five  covariants/invariants are given as follows:
\begin{enumerate}
\item Three covariants
\begin{gather}
\begin{split}\label{eq:ABCgammas}
(\gamma^{A})_{A_{1}A_{2}}&=a\indices{_{A_{1}}^{BC}}a_{A_{2}BC}, \\
(\gamma^{B})_{B_{1}B_{2}}&=a\indices{^{A}_{B_{1}}^{C}}a_{AB_{2}C}, \\
(\gamma^{C})_{C_{1}C_{2}}&=a\indices{^{AB}_{C_{1}}}a_{ABC_{2}},
\end{split}
\end{gather}
transforming, respectively, as a $\rep{(3,1,1)}$, $\rep{(1,3,1)}$, and $\rep{(1,1,3)}$  under $SL_A(2, \mathds{C}) \times SL_B(2, \mathds{C}) \times SL_C(2, \mathds{C})$.
\item One covariant $T_{ABC}$ transforming as a $\bf {(2,2,2)}$ under $[SL(2, \mathds{C})]^3$, which may be written in one of three equivalent forms
\begin{equation}\label{eq:Tofgamma}
\begin{split}
T_{ABC}&=(\gamma^{A})_{AA'}a\indices{^{A'}_{BC}}\\
T_{ABC}&=(\gamma^{B})_{BB'}a\indices{_{A}^{B'}_{C}}\\
T_{ABC}&=(\gamma^{C})_{CC'}a\indices{_{AB}^{C'}}.
\end{split}
\end{equation}
\item Cayley's hyperdeterminant $\Det a_{ABC}$
\cite{Cayley:1845, Miyake:2002, Miyake:2003}, the unique quartic $[SL(2, \mathds{C})]^3$ invariant,
where
\begin{equation}\label{eq:hyperequalsdetgamma}
\Det a_{ABC}=-\det \gamma^{A}=-\det \gamma^{B}=-\det \gamma^{C}.
\end{equation}
\end{enumerate}
The entanglement classification as determined by these covariants/invariants is summarized in \autoref{tab:merge}.
\begin{table}
\caption{The entanglement classification of three qubits.\label{tab:merge}}
\begin{ruledtabular}
\begin{tabular*}{\textwidth}{@{\extracolsep{\fill}}cccM{c}M{c}c}
& Class       & & \text{Vanishing}             & \text{Nonvanishing} & \\
\hline
& $A$-$B$-$C$ & & \gamma^A, \gamma^B, \gamma^C & a_{ABC}              & \\
& $A$-$BC$    & & \gamma^B, \gamma^C           & \gamma^A             & \\
& $B$-$CA$    & & \gamma^A, \gamma^C           & \gamma^B             & \\
& $C$-$AB$    & & \gamma^A, \gamma^B           & \gamma^C             & \\
& W           & & \Det a_{ABC}                 & T_{ABC}              & \\
& GHZ         & & \cdots                       & \Det a_{ABC}         & \\
\end{tabular*}
\end{ruledtabular}
\end{table}

There are six independent $[SU(2)]^3$ pure state invariants \cite{Sudbery:2001}: the norm, the three local entropies $4\det \rho_A$, $4\det \rho_B$, $4\det \rho_C$, the Kempe invariant \cite{Kempe:1999vk}, and finally the all important 3-tangle $\tau_{ABC}$ \cite{Coffman:1999jd},
\begin{equation}
\tau_{ABC} = 4|\Det a_{ABC}|.
\label{eq:3tangle}
\end{equation}
The 3-tangle is maximized, $\tau_{ABC}=1$, by the GHZ state:
\begin{equation}
\ket{\Psi}_{\textrm{GHZ}}= \tfrac{1}{\sqrt{2}}(\ket{000}+\ket{111}).
\end{equation}

\section{Superqubits}
\label{sec:superqubits}

\subsection{Super Hilbert space and $uOSp(1|2)$}
\label{sec:superH}

\subsubsection{The dual space}

With one important difference, explained below, our definition of a super Hilbert space follows that of DeWitt \cite{DeWitt:1984}.  We define a super Hilbert space to be a supervector space $\mathcal{H}$ equipped with an injection to its dual space $\mathcal{H}^\ddag$,
\begin{equation}
\begin{split}
{}^\ddag:\mathcal{H}&\to\mathcal{H}^\ddag,\\
          \ket{\psi}&\mapsto (\ket{\psi})^\ddag:=\bra{\psi}.
\end{split}
\end{equation}

Details of even and odd Grassmann numbers and supervectors may be found in \hyperref[sec:toolkit]{Appendix~\ref*{sec:toolkit}}.  A basis in which all basis vectors are pure even or odd is said to be pure. Such a basis may always be found \cite{DeWitt:1984}.

The map ${}^\ddag:\mathcal{H}\to\mathcal{H}^\ddag$ defines an inner product $\braket{\psi}{\phi}$ and satisfies the following axioms:
\begin{enumerate}
\item $^\ddag$ sends pure bosonic (fermionic) supervectors in $\mathcal{H}$ into bosonic (fermionic) supervectors in $\mathcal{H}^\ddag$.
	
\item $^\ddag$ is linear
\begin{equation}
(\ket{\psi}+ \ket{\phi})^\ddag=\bra{\psi}+ \bra{\phi}.
\end{equation}
	
\item For pure even/odd $\alpha$ and $\ket{\psi}$
\begin{equation}\label{eq:ketadjoint}
(\ket{\psi}\alpha)^\ddag=(-)^{\alpha\psi}\alpha^\#\bra{\psi}
\end{equation}
and
\begin{equation}\label{eq:braadjoint}
(\alpha\bra{\psi})^\ddag=(-)^{\psi+\alpha\psi}\ket{\psi}\alpha^\#,
\end{equation}
where ${}^\#$ is the superstar introduced in \hyperref[sec:toolkit]{Appendix~\ref*{sec:toolkit}}. In particular,
\begin{equation}\label{eq:braadjoint1}
\ket{\psi}^{\ddag\ddag}=(-)^{\psi}\ket{\psi}.
\end{equation}
Note, an $\alpha$ (or $\psi$ and the like) appearing in the exponent of $(-)$ is shorthand for its grade, $\deg(\alpha)$, which takes the value $0$ or $1$ according to whether $\alpha$ is even or odd. The impure case follows from the linearity of $^{\ddag}$.
\end{enumerate}

In a pure even/odd orthonormal basis $\{\ket{i}\}$ we adopt the following convention:
\begin{equation}
\ket{\psi}=\ket{i}\psi_i
\end{equation}
so that for pure even/odd $\psi$ \eqref{eq:ketadjoint} and \eqref{eq:braadjoint} imply
\begin{equation}
\begin{split}
(\ket{i}\psi_i)^\ddag &=(-)^{\psi_ii}\psi_{i}^{\#}\bra{i}=(-)^{i+i\psi}\psi_{i}^{\#}\bra{i}\\
((-)^{i+i\psi}\psi_{i}^{\#}\bra{i})^\ddag &=(-)^{\psi}\ket{i}\psi_i
\end{split}
\end{equation}
where we have used $\deg(\psi_i)=\deg(i)+\deg(\psi)$. This is consistent with \eqref{eq:scalarmultiplication}.

\subsubsection{Inner product}
	
For all pure even/odd $\ket{\psi}, \ket{\phi}\in \mathcal{H}$ the inner product $\braket{\psi}{\phi}$ satisfies
\begin{equation}
\braket{\psi}{\phi}^\#=(-)^{\psi+\psi\phi}\braket{\phi}{\psi}.
\end{equation}
Consequently,
\begin{equation}
\braket{\psi}{\phi}^{\#\#}=(-)^{\psi+\phi}\braket{\phi}{\psi},
\end{equation}
as would be expected of a pure even/odd Grassmann number since $\deg(\braket{\phi}{\psi})=\deg(\psi)+\deg(\phi)$. In a pure even/odd orthonormal basis we find
\begin{equation}
\braket{\phi}{\psi}=(-)^{i+i\phi}\phi_{i}^{\#}\psi_i.
\end{equation}
In using the superstar we depart from the formalism presented in \cite{DeWitt:1984}, which uses the ordinary star.  A comparison of the star and superstar may be found in \hyperref[sec:toolkit]{Appendix~\ref*{sec:toolkit}} . The use of the superstar anticipates the implementation of $uOSp(1|2)$ as the compact subgroup of $OSp(1|2)$ as will be explained in \autoref{sec:superSLOCC}.
	
\subsubsection{Linear superoperators and the superadjoint}

A linear superoperator $A:\mathcal{H}\rightarrow\mathcal{H}$ is required to satisfy the following properties:
\begin{enumerate}
\item $A(\ket{\psi}+\ket{\phi})=A\ket{\psi}+A\ket{\phi},$
\item $A(\ket{\psi}\alpha)=(A\ket{\psi})\alpha.$
\end{enumerate}
Linear superoperators may be combined using
\begin{enumerate}
\item $(A+B)\ket{\psi}=A\ket{\psi}+B\ket{\psi},$
\item $(AB)\ket{\psi}=A(B\ket{\psi}).$
\end{enumerate}
A linear superoperator is said to be pure even (odd) if it takes pure even supervectors into pure even (odd) supervectors and pure odd supervectors into pure odd (even) supervectors.

The superadjoint of a pure even/odd linear superoperator is defined through
\begin{equation}\label{eq:opadjoint1}
(A\ket{\phi})^\ddag=(-)^{\phi A}\bra{\phi}A^\ddag.
\end{equation}
This is in fact equivalent to
\begin{equation}\label{eq:opadjoint2}
\bra{\phi}A^\ddag\ket{\psi}=(-)^{\psi+\phi\psi +(\phi+\psi)A}\bra{\psi}A\ket{\phi}^\#,
\end{equation}
which is the natural supersymmetric generalization of the conventional definition of the adjoint. This equivalence may be established by simply inserting the identity operator, $\mathds{1}=\ketbra{i}{i}$, in \eqref{eq:opadjoint1},
\begin{widetext}
\begin{equation}
\begin{array}{cr@{}l@{\ =\ }r@{}l}
&(\ketbra{i}{i}&A\ket{\phi})^\ddag &(-)^{\phi A}\bra{\phi}&A^\ddag\ketbra{i}{i}\\
\Rightarrow&(-)^{i(i+A+\phi)}\bra{i}&A\ket{\phi}^\#\bra{i} &(-)^{\phi A}\bra{\phi}&A^\ddag\ketbra{i}{i}\\
\Rightarrow&(-)^{i+i\phi+(i+\phi)A}\bra{i}&A\ket{\phi}^\# &\bra{\phi}&A^\ddag\ket{i}\\
\Rightarrow&\sum_{i}(-)^{i+i\phi+(i+\phi)A}\bra{i}&A\ket{\phi}^\#\psi_i &\sum_i\bra{\phi}&A^\ddag\ket{i}\psi_i\\
\Rightarrow&\sum_{i}(-)^{i+i\phi+(i+\phi)A+\psi_i(i+A+\phi)+\psi_i}(\psi_{i}^{\#}\bra{i}&A\ket{\phi})^\# &\sum_i\bra{\phi}&A^\ddag\ket{i}\psi_i\\
\Rightarrow&(-)^{\psi+\phi\psi+(\psi+\phi)A}\bra{\psi}&A\ket{\phi}^\# &\bra{\phi}&A^\ddag\ket{\psi},
\end{array}
\end{equation}
\end{widetext}
where we have defined $\ket{\psi}=\ket{i}\psi_i$ and used $\deg(\psi)=\deg(\psi_i)+\deg(i)$. The converse implication follows from a similar treatment, which we omit.
From \eqref{eq:opadjoint2} we also have
\begin{equation}
(\bra{\phi}A)^\ddag=(-)^{\phi+\phi A}A^\ddag\ket{\phi}.
\end{equation}
Moreover,		
\begin{equation}
A^{\ddag\ddag}=(-)^AA,
\end{equation}
which is consistent with the properties of supermatrices and the supermatrix superadjoint given in \hyperref[sec:toolkit]{Appendix~\ref*{sec:toolkit}}.

In a pure even/odd orthonormal basis the supermatrix representation of a linear operator $A$ is given by
\begin{equation}
A_{ij}:=\bra{i}A\ket{j}.
\end{equation}
In particular, \eqref{eq:opadjoint2} implies that the component form of the adjoint is given by
\begin{equation}(A^\ddag)_{ij}=(-)^{j+ij+(i+j)A}A_{ji}^{\#},\end{equation}
where an index in the exponent of $(-)$ is understood to take the value $0$ or $1$ according to whether it corresponds to an  even or odd basis vector. This is just the conventional supermatrix superadjoint used to define $uOSp(1|2)$ in \autoref{sec:superSLOCC}.
			
For any linear operator of the form $\ket{\psi}\bra{\phi}$ one obtains
\begin{equation}\label{eq:butteradjoint}
(\ket{\psi}\bra{\phi})^\ddag=(-)^{\phi+\phi\psi}\ket{\phi}\bra{\psi}.
\end{equation}
For pure even/odd $\ket{\psi}$ the butterfly operator $\ket{\psi}\bra{\psi}$ is manifestly self-adjoint.
			
The inner product is invariant under the action of all even operators satisfying the superunitary condition
\begin{equation}
A^\ddag A=\mathds{1}, \qquad A^{\ddag}_{ij}A_{jk}=\delta_{ik}.
\end{equation}
Let $\ket{\psi}$ be a pure even/odd supervector and
\begin{equation}
\ket{\tilde{\psi}}=A\ket{\psi}.
\end{equation}
Then, in a pure orthonormal basis $\{\ket{i}\}$
\begin{equation}
\begin{split}
\tilde{\psi}_i &=\braket{i}{\tilde{\psi}}\\
               &=\bra{i}A\ket{j}\psi_j\\
               &=A_{ij}\psi_j.
\end{split}
\end{equation}
Hence, for pure even/odd supervectors $\ket{\phi}$ and $\ket{\psi}$ and even $A$ the transformed inner product is given by
\begin{equation}
\begin{split}
\braket{\tilde{\phi}}{\tilde{\psi}} &=(-)^{i+i\tilde{\phi}}\tilde{\phi}_{i}^\#\tilde{\psi}_i\\
                                    &=(-)^{i+i\phi}(A_{ij}\phi_{j})^\#A_{ik}\psi_k\\
                                    &=(-)^{i+i\phi+(j+\phi)(i+j)}\phi_{j}^{\#}A_{ij}^{\#}A_{ik}\psi_k\\
                                    &=(-)^{i+i\phi+(j+\phi)(i+j)}\phi_{j}^{\#}(-)^{i+ij}A_{ji}^{st\#}A_{ik}\psi_k\\
                                    &=(-)^{(j+j\phi )}\phi_{j}^{\#}A_{ji}^{\ddag}A_{ik}\psi_k\\
                                    &=(-)^{(j+j\phi )}\phi_{j}^{\#}\psi_j\\
                                    &=\braket{\phi}{\psi}
\end{split}
\end{equation}
where we have used $\deg(A_{ij})=\deg(i)+\deg(j)$.

\subsubsection{Physical states}
		
For all $\ket{\psi}\in \mathcal{H}$
	\begin{equation}
	\braket{\psi}{\psi}_{\mathcal{B}}\geq 0.
	\end{equation}	
Here $z_{\mathcal{B}}\in\mathds{C}$ denotes the purely complex number component of the Grassmann number $z$ and is referred to as the \emph{body}, a terminology introduced in  \cite{DeWitt:1984}. The \emph{soul} of $z$, denoted $z_\mathcal{S}$, is the purely Grassmannian component. Any Grassmann number may be decomposed into body and soul, $z=z_{\mathcal{B}}+z_\mathcal{S}$.

A Grassmann number has an inverse iff it has a nonvanishing body. Consequently, a state $\ket{\psi}$ is normalizable iff $\braket{\psi}{\psi}_{\mathcal{B}}> 0$. The state may then be normalized,
\begin{equation}
\ket{\hat{\psi}}=N_\psi\ket{\psi},\quad N_\psi=\braket{\psi}{\psi}^{-1/2},
\end{equation}
where $N_\psi$ is given by the general definition of an analytic function $f$ on the space of Grassmann numbers \eqref{eq:analyticgrassmann}. Explicitly,
\begin{equation}\label{eq:grassfuncsqrt}
\braket{\psi}{\psi}^{-1/2}=\sum^{\infty}_{k=0}\frac{1}{k!2^k}\prod_{j=0}^{k}(1-2j)\braket{\psi}{\psi}_{\mathcal{B}}^{-\frac{2k+1}{2}}\braket{\psi}{\psi}_{\mathcal{S}}^{k}.
\end{equation}

Motivated by the above considerations a state $\ket{\psi}$ is said to be \emph{physical} iff $\braket{\psi}{\psi}_{\mathcal{B}}> 0$.  We restrict our attention to physical states throughout.

\subsection{Super LOCC and SLOCC }\label{sec:superSLOCC}

We promote the conventional SLOCC equivalence group $SL(2,\mathds{C})$ to its minimal supersymmetric extension $OSp(1|2)$ \cite{VanProeyen:1999ni,Frappat:2000}. The orthosymplectic superalgebras and $OSp(1|2)$, in particular, are described in \hyperref[sec:osp]{Appendix~\ref*{sec:osp}}.

The three even elements $P_{A_1A_2}$ form an $\mathfrak{sl}(2)$ subalgebra generating the bosonic SLOCC equivalence group, under which  $Q_A$ transforms as a spinor.

The supersymmetric generalization of the conventional group of local unitaries is given by $uOSp(1|2)$, a compact subgroup of $OSp(1|2)$ \cite{Berezin:1981, Frappat:2000}. It has a supermatrix representation as the subset of  $OSp(1|2)$ supermatrices satisfying the additional superunitary condition
\begin{equation}
M^\ddag M =\mathds{1},
\end{equation}
where $^\ddag$ is the superadjoint given by
\begin{equation}
M^\ddag= (M^{st})^\#.
\end{equation}
The $uOSp(1|2)$ algebra is given by
\begin{equation}
\mathfrak{uosp}(1|2):=\{X\in\mathfrak{osp}(1|2)|X^\ddag=-X\}.
\end{equation}
An arbitrary element $X\in\mathfrak{uosp}(1|2)$ may be written as
\begin{equation}
X=\xi_i A_i + \eta^\# Q_0 + \eta Q_1,
\end{equation}
where $\xi_i$ and $\eta$ are pure even/odd Grassmann numbers respectively and
\begin{equation}
\begin{gathered}
A_1=\tfrac{i}{2}(P_{00}- P_{11}), \qquad A_2=\half(P_{00}+ P_{11}),\\
 A_3=iP_{01},\\
Q_{A}^{\ddag}=\eps_{AA'}Q_{A'}, \qquad A_{i}^{\ddag}=-A_{i}.
\end{gathered}
\end{equation}

\subsection{One superqubit}

The one-superqubit system (Alice) is described by the state
\begin{equation}
\ket{\Psi} = \ket{A}a_{A} +\ket{\bp}a_{\bp},
\end{equation}
where ${ a}_{A}$ is commuting with $A=0,1$ and $a_{\bp}$ is anticommuting. That is to say, the state vector is promoted to a supervector. The super Hilbert space has dimension 3, two ``bosons'', and one ``fermion''.  In more compact notation we may write
\begin{equation}
\ket{\Psi} = \ket{X}a_{X},
\end{equation}
where $X=(A,\bp)$.

The super SLOCC  equivalence group for a single qubit is $OSp(1|2)_{A}$. Under the  $SL(2)_{A}$ subgroup $a_{A}$ transforms as a \textbf{2} while $a_{\bp}$ is a singlet as shown in \autoref{tab:action1}. The super LOCC entanglement equivalence group, i.e. the group of local unitaries, is given by $uOSp(1|2)_{A}$, the unitary subgroup of $OSp(1|2)_{A}$.
\begin{table}
\caption{The action of the $\mathfrak{osp}(1|2)$ generators on the superqubit fields.\label{tab:action1}}
\begin{ruledtabular}
\begin{tabular}{c*{5}{M{c}}c}
& \multirow{2}{*}{Generator} && \multicolumn{2}{c}{Field acted upon} && \\
\cline{3-6}
&                            && a_{A_3}                  & a_{\bp}   && \\
\hline
& P_{A_1A_2}                 && \eps_{(A_1|A_3}a_{|A_2)} & 0         && \\
& 2Q_{A_1}                    && \eps_{A_1A_3}a_{\bp}     & a_{A_1}   &&
\end{tabular}
\end{ruledtabular}
\end{table}

The norm squared $\braket{\Psi}{\Psi}$ is given by
\begin{equation}
\langle \Psi | \Psi \rangle = \delta^{A_1A_2} a_{A_1}^{\#}a_{A_2} - a_{\bp}^{\#}a_{\bp},
\end{equation}
where $\bra{\Psi}= (\ket{\Psi})^\ddag$ and $\braket{\Psi}{\Psi}$ is the conventional inner product that is manifestly $uOSp(1|2)$ invariant. The one-superqubit state may then be normalized.

As explained in \hyperref[sec:toolkit]{Appendix~\ref*{sec:toolkit}} the $n$-superqubit Hilbert space is defined over a $2^{n+1}$-dimensional Grassmann algebra for which $z^{2n+1}_\mathcal{S}=0$ for all $z$. So \eqref{eq:grassfuncsqrt} terminates after a finite number of terms:
\begin{equation}\label{eq:grassfuncsqrtnqubit} \braket{\Psi}{\Psi}^{-1/2}=\sum^{n}_{k=0}\frac{1}{k!2^k}\prod_{j=0}^{k}(1-2j)\braket{\Psi}{\Psi}_{\mathcal{B}}^{-\frac{2k+1}{2}}\braket{\Psi}{\Psi}_{\mathcal{S}}^{k},
\end{equation}
where the sum only runs to $n$ since the bracket $\braket{\Psi}{\Psi}_{\mathcal{S}}$ is at least quadratic in Grassmann variables.
For one superqubit, with $a_{A}$ pure body, this gives
\begin{equation}
\begin{split}
\langle \Psi | \Psi \rangle^{-1/2}= \quad&(\delta^{A_1A_2}a^{*}_{A_1}a_{A_2})^{-1/2}\\
+\half &(\delta^{A_1A_2}a^{*}_{A_1}a_{A_2})^{-3/2}a_{\bp}^{\#}a_{\bp}
\end{split}
\end{equation}
so the normalized wave function $\ket{\hat \Psi}$, for which $\braket{\hat \Psi}{\hat \Psi}=1$, is
\begin{equation}
| \hat \Psi \rangle=| A \rangle{\hat a}_A + | \bp \rangle{\hat a}_{\bp}
\end{equation}
where
\begin{equation}
\begin{split}
{\hat a}_A&= a_A[(\delta^{A_1A_2}a^{*}_{A_1}a_{A_2})^{-1/2}\\
&\qquad+\half (\delta^{A_1A_2}a^{*}_{A_1}a_{A_2})^{-3/2}a_{\bp}^{\#}a_{\bp}],\\
{\hat a}_{\bp}&= { a}_{\bp}(\delta^{A_1A_2}a^{*}_{A_1}a_{A_2})^{-1/2}.
\end{split}
\end{equation}

The one-superqubit density matrix is given by
\begin{equation}
\begin{split}
\rho&:=\ketbra{\Psi}{\Psi}= (-)^{X_2}\ket{X_1}a_{X_1}a_{X_2}^{\#}\bra{X_2}\\
&= \ket{A_1}a_{A_1}a_{A_2}^{\#}\bra{A_2}- \ket{A_1}a_{A_1}a_{\bp}^{\#}\bra{\bp }\\
&\phantom{=} +\ket{\bp}a_\bp a_{A_2}^{\#}\bra{A_2}-\ket{\bp}a_{\bp}a_{\bp}^{\#}\bra{\bp}.
\end{split}
\end{equation}
Alternatively, in components, we may write
\begin{equation}
\begin{split}
\rho_{X_1X_2}&=\bra{X_1}\rho\ket{X_2}\\
&=(-)^{X_2}a_{X_1} a_{X_2}^\#.
\end{split}
\end{equation}
The density matrix is self-superadjoint,
\begin{equation}
\begin{split}
\rho^{\ddag}_{X_1X_2}&=(\rho^{st}_{X_1X_2})^{\#}\\
&=(-)^{X_2+X_1X_2}\rho_{X_2X_1}^\#\\
    &=(-)^{X_2+X_1X_2}(-)^{X_1}a_{X_2}^\#a_{X_1}^{\#\#}\\
      &=(-)^{X_2}a_{X_1}a_{X_2}^\#\\
      &=\rho_{X_1X_2}.
\end{split}
\end{equation}
The norm squared is then given by the supertrace
\begin{equation}
\begin{split}
\str(\rho)&= (-)^{X_1}\delta^{X_1X_2}\bra{X_1}\rho\ket{X_2}\\
&=\sum_X a_{X}a_{X}^\#\\
&=\sum_X (-)^Xa_{X}^\#a_{X}\\
&=\braket{\Psi}{\Psi}
\end{split}
\end{equation}
as one would expect.

Unnormalized pure state super density matrices satisfy $\rho^2 = \str(\rho)\rho$,
\begin{equation}\label{eq:SqubitMixedState}
\begin{split}
\rho^2&=(-)^{X_2}a_{X_1} a_{X_2}^\#\delta^{X_2X_3}(-)^{X_4}a_{X_3} a_{X_4}^\#\\
      &= \delta^{X_2X_3}a_{X_2}a_{X_3}^\#(-)^{X_4}a_{X_1} a_{X_4}^\#\\
      &= \str(\rho)\rho,
\end{split}
\end{equation}
the appropriate supersymmetric version of  the conventional pure state density matrix condition \eqref{eq:densitycondition}.

\subsection{Two superqubits}

The two-superqubit system (Alice and Bob) is described by the state
\begin{equation}
\ket{\Psi} = \ket{AB}a_{AB} +\ket{A\bp}a_{A\bp}+\ket{\bp B}a_{\bp B} +\ket{\bp\bp}a_{\bp\bp}
\end{equation}
where $a_{AB}$ is commuting, $a_{A\bp}$ and $a_{\bp B}$ are anticommuting and $a_{\bp\bp}$ is commuting. The super Hilbert space has dimension 9: 5 bosons and 4 fermions. The super SLOCC group for two superqubits is $OSp(1|2)_{A} \times OSp(1|2)_{B}$. Under the  $SL(2)_{A} \times SL(2)_{B}$ subgroup $a_{AB}$ transforms as a $\rep{(2,2)}$, $a_{A\bp}$ as a $\rep{(2,1)}$, $a_{\bp B}$ as a $\rep{(1,2)}$ and $a_{\bp\bp}$ as a $\rep{(1,1)}$ as summarized in \autoref{tab:action2}.
\begin{table*}
\caption{The action of the $\mathfrak{osp}(1|2)\oplus \mathfrak{osp}(1|2)$ generators on the 2-superqubit fields.\label{tab:action2}}
\begin{ruledtabular}
\begin{tabular}{c*{9}{M{c}}c}
& \multirow{3}{*}{Generator} && \multicolumn{6}{c}{Field acted upon}                                                                      && \\
&                            && \multicolumn{2}{c}{Bosons}                 &&& \multicolumn{2}{c}{Fermions}                               && \\
\cline{3-6}\cline{7-10}
&                            && a_{A_3B_3}                  & a_{\bp\bp}   &&& a_{A_3\bp}                  & a_{\bp B_3}                  && \\
\hline
& P_{A_1A_2}                 && \eps_{(A_1|A_3}a_{|A_2)B_3} & 0            &&& \eps_{(A_1|A_3}a_{|A_2)\bp} & 0                            && \\
& P_{B_1B_2}                 && \eps_{(B_1|B_3}a_{A_3|B_2)} & 0            &&& 0                           & \eps_{(B_1|B_3}a_{\bp|B_2)}  && \\
& 2Q_{A_1}                    && \eps_{A_1A_3}a_{\bp B_3}    & a_{A_1\bp}   &&& \eps_{A_1A_3}a_{\bp\bp }    & a_{A_1B_3}                   && \\
& 2Q_{B_1}                    && \eps_{B_1B_3}a_{A_3\bp}     & -a_{\bp B_1} &&& a_{A_3B_1}                  & -\eps_{B_1B_3}a_{\bp\bp}     &&
\end{tabular}
\end{ruledtabular}
\end{table*}
The coefficients may also be assembled into a  $(2|1)\times(2|1)$ supermatrix
\begin{figure}
\centering
\includegraphics[width=.35\textwidth]{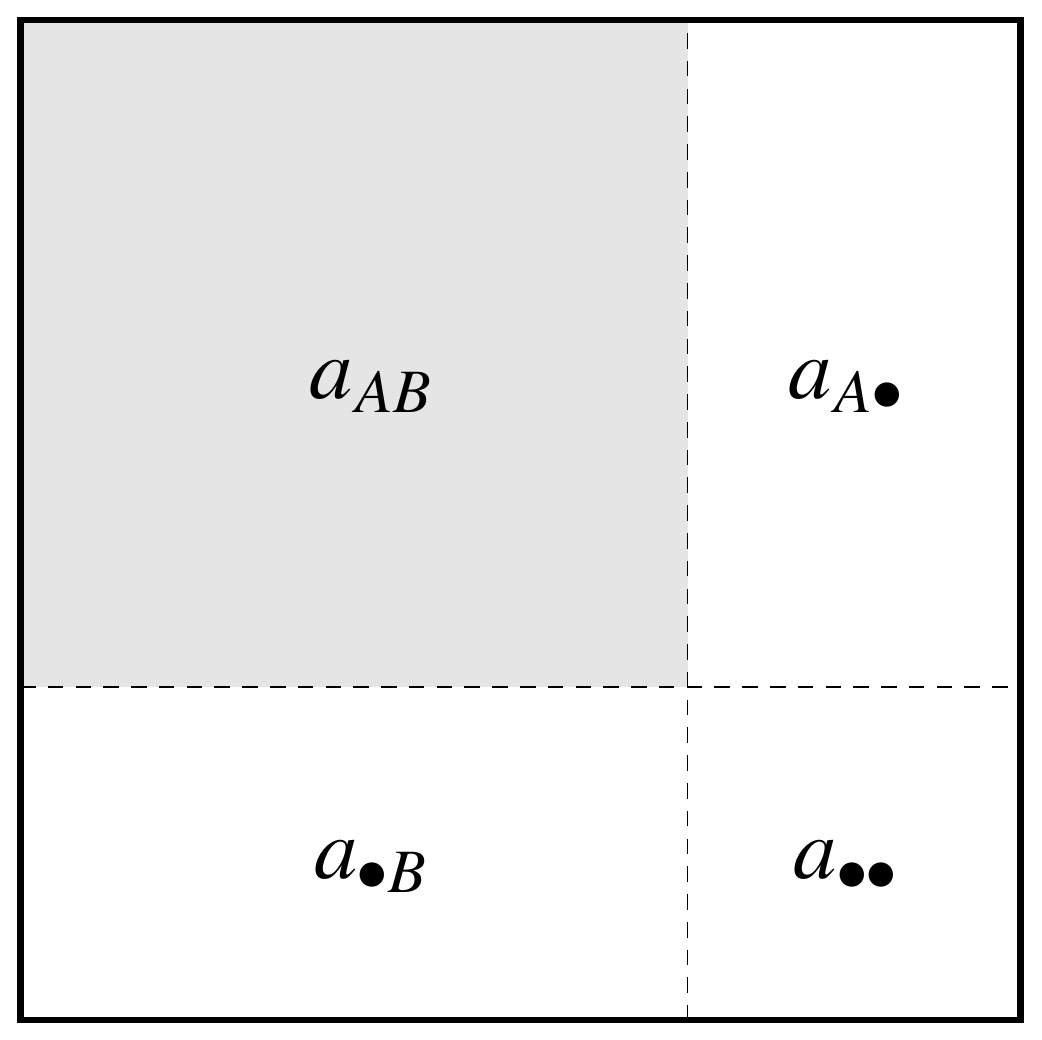}
\caption{The $3 \times 3$ square supermatrix\label{fig:supermatrix}}
\end{figure}
\begin{equation}
\braket{XY}{\Psi}=a_{XY} = \begin{pmatrix}a_{AB}&\vrule&a_{A\bp}\\ \hline a_{\bp B}&\vrule&a_{\bp\bp}\end{pmatrix}.
\end{equation}
See \autoref{fig:supermatrix}.

The norm squared $\braket{\Psi}{\Psi}$ is given by
\begin{equation}
\begin{split}
\langle \Psi | \Psi \rangle &=(-)^{X_1+Y_1}\delta^{X_1X_2}\delta^{Y_1Y_2}a_{X_1Y_1}^{\#}a_{X_2Y_2}\\
&=\delta^{A_1A_2} \delta^{B_1B_2}a_{A_1B_1}^{\#}a_{A_2B_2} \\
&\phantom{=}- \delta^{A_1A_2}a_{ A_1\bp}^{\#}a_{A_1\bp}- \delta^{B_1B_2}a_{\bp B_1}^{\#}a_{\bp B_1}\\
&\phantom{=}+ a_{\bp\bp}^{\#}a_{\bp\bp},
\end{split}
\end{equation}
where $\bra{\Psi}= (\ket{\Psi})^\ddag$ and $\langle \Psi | \Psi \rangle$ is the conventional inner product taht is manifestly $uOSp(1|2)_A\times uOSp(1|2)_B$ invariant.

The two-superqubit density matrix is given by
\begin{equation}
\begin{split}
\rho&=\ketbra{\Psi}{\Psi}\\
&= (-)^{X_2+Y_2}\ket{X_1Y_1}a_{X_1Y_1}a_{X_2Y_2}^\#\bra{X_2Y_2}.
\end{split}
\end{equation}
The reduced density matrices for Alice and Bob are given by the partial supertraces:
\begin{subequations}
\begin{gather}
\begin{split}
\rho_A&=\sum_Y(-)^Y \bra{Y}\rho \ket{Y}\\
      &=\sum_Y(-)^{X_2} \ket{X_1}a_{X_1Y}a_{X_2Y}^\#\bra{X_2},
\end{split}\\
\begin{split}
\rho_B&=\sum_X(-)^X \bra{X}\rho \ket{X}\\
      &=\sum_X(-)^{Y_2} \ket{Y_1}a_{XY_1}a_{XY_2}^\#\bra{Y_2}.
\end{split}
\end{gather}
\end{subequations}

In component form the reduced density matrices are given by
\begin{equation}
\begin{split}
(\rho_A)_{X_1X_2}&=\sum_Y(-)^{X_2} a_{X_1Y}a_{X_2Y}^\#,\\
(\rho_B)_{Y_1Y_2}&=\sum_X(-)^{Y_2} a_{XY_1}a_{XY_2}^\#,
\end{split}
\end{equation}
and
\begin{equation}
\str{\rho_A}=\str{\rho_B}=\braket{\Psi}{\Psi}.
\end{equation}

\subsection{Three superqubits}

The three-superqubit system (Alice, Bob, and Charlie) is described by the state
\begin{equation}
\begin{gathered}
\ket{\Psi} = \ket{ABC}a_{ABC}\\
+\ket{AB\bp}a_{AB\bp}+\ket{A\bp C}a_{A\bp C}+\ket{\bp BC}a_{\bp BC}\\
+\ket{A\bp\bp}a_{A\bp\bp}+\ket{\bp B \bp}a_{\bp B \bp}+\ket{\bp \bp C}a_{\bp \bp C}\\
+\ket{\bp \bp \bp }a_{\bp \bp \bp}
\end{gathered}
\end{equation}
where $a_{AB}$ is commuting, $a_{AB\bp}$  $a_{A\bp C}$ $a_{\bp BC}$ are anticommuting, $a_{A\bp\bp}$ $a_{\bp B \bp}$ $a_{\bp \bp C}$ are commuting and $a_{\bp \bp\bp}$ is anticommuting. The super Hilbert space has dimension 27: 14 bosons and 13 fermions. The super SLOCC group for three superqubits is $OSp(1|2)_{A} \times OSp(1|2)_{B} \times OSp(1|2)_{C}  $. Under the  $SL(2)_{A} \times SL(2)_{B} \times SL(2)_{C}$ subgroup $a_{ABC}$ transforms as a $\rep{(2,2,2)}$, $a_{AB\bp}$ as a $\rep{(2,1,1)}$, $a_{A \bp C}$ as a $\rep{(2,1,2)}$, $a_{ \bp BC}$ as a $\rep{(1,2,2)}$, $a_{A \bp\bp}$ as a $\rep{(2,1,1)}$, $a_{ \bp B \bp}$ as a $\rep{(1,2,1)}$, $a_{\bp\bp C}$ as a $\rep{(1,1,2)}$, and $a_{\bp \bp \bp}$ as a $\rep{(1,1,1)}$ as summarized in \autoref{tab:action3}.The coefficients may also be assembled into a  $(2|1)\times(2|1)\times(2|1)$ superhypermatrix
\begin{figure}
\centering
\includegraphics[width=.4\textwidth]{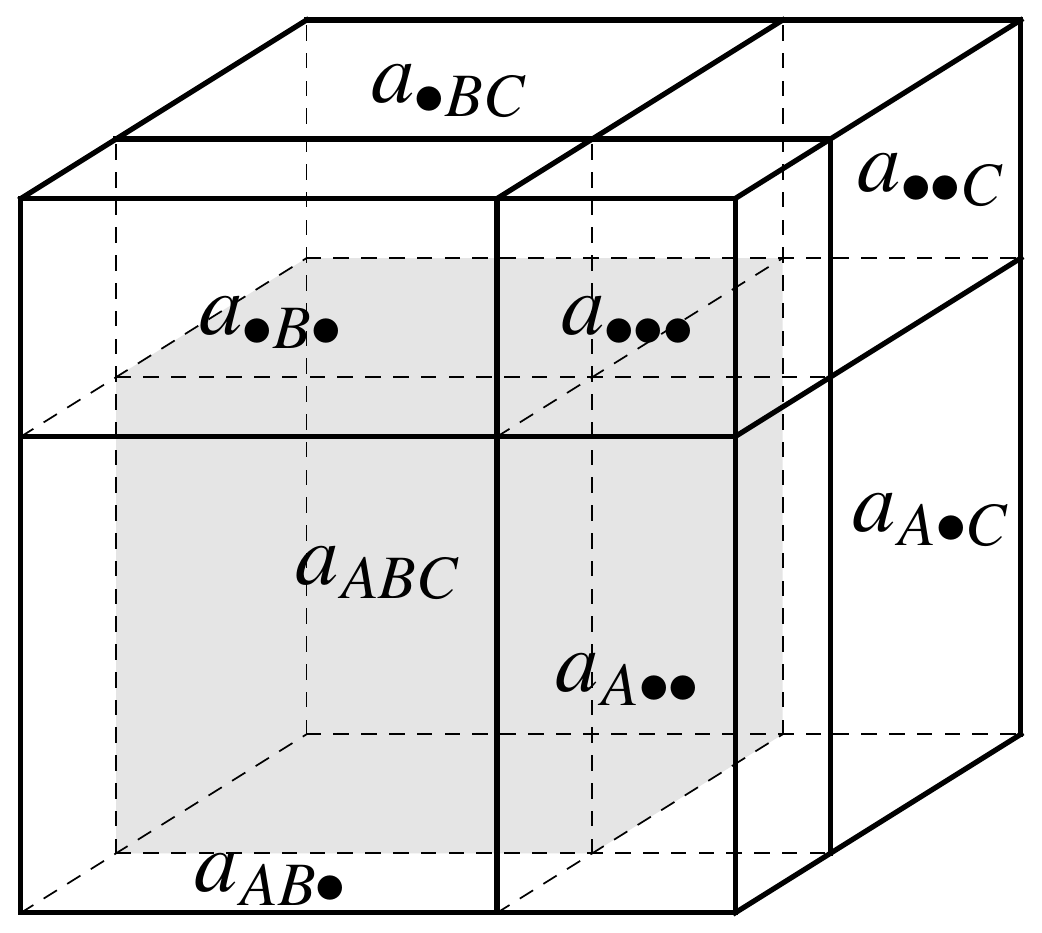}
\caption{The $3 \times 3 \times 3$ cubic superhypermatrix}
\label{fig:superhypermatrix}
\end{figure}
\begin{equation}
\braket{XYZ}{\Psi}=a_{XYZ}.
\end{equation}
See \autoref{fig:superhypermatrix}.

\begin{table*}
\caption{The action of the $\mathfrak{osp}(1|2)\oplus \mathfrak{osp}(1|2)\oplus \mathfrak{osp}(1|2)$ generators on the 3-superqubit fields.\label{tab:action3}}
\begin{ruledtabular}
\begin{tabular}{cM{c}c*{4}{M{c}}cc}
& \multirow{2}{*}{Generator} & & \multicolumn{4}{c}{Bosons acted upon}  & &\\

\cline{3-7}
&                            & & a_{A_3B_3C_3}                  & a_{A_3 \bp\bp}                 & a_{\bp B_3\bp}                 & a_{\bp\bp C_3}                  & & \\
\hline
& P_{A_1A_2}                 & & \eps_{(A_1|A_3}a_{|A_2)B_3C_3} & \eps_{(A_1|A_3}a_{|A_2)\bp\bp} & 0                              & 0                               & & \\
& P_{B_1B_2}                 & & \eps_{(B_1|B_3}a_{A_3|B_3)C_2} & 0                              & \eps_{(B_1|B_3}a_{\bp|A_2)\bp} & 0                                 & & \\
& P_{C_1C_2}                 & & \eps_{(C_1|C_3}a_{A_3B_3|C_2)} & 0                              & 0                              & \eps_{(C_1|C_3}a_{\bp\bp|C_2)}      & & \\
\hline
& 2Q_{A_1}                   & & \eps_{A_1A_3}a_{\bp B_3C_3}    & \eps_{A_1A_3}a_{\bp\bp\bp}     & a_{A_1B_3\bp }                 & a_{A_1\bp C_3}                 & & \\

& 2Q_{B_1}                   & & \eps_{B_1B_3}a_{A_3\bp C_3}    & a_{A_3B_1 \bp}                & -\eps_{B_1B_3}a_{\bp\bp\bp}     & -a_{\bp B_1C_3}                  & & \\

& 2Q_{C_1}                   & & \eps_{C_1C_3}a_{A_3B_3 \bp}    & -a_{A_3\bp C_1}                 & -a_{\bp B_3C_1}                & \eps_{C_1C_3}a_{\bp\bp\bp}     & &\\

\hline

&                            & &  \multicolumn{4}{c}{Fermions acted upon} 		& &\\
\cline{3-7}
&                            & & a_{A_3B_3\bp}                  & a_{A_3\bp C_3}                  & a_{\bp B_3C_3}                  & a_{\bp\bp\bp}  & & \\
\hline
& P_{A_1A_2}                 & & \eps_{(A_1|A_3}a_{|A_2)B_3\bp} & \eps_{(A_1|A_3}a_{|A_2)\bp C_3} & 0                               & 0              & & \\
& P_{B_1B_2}                 & & \eps_{(B_1|B_3}a_{A_3|B_3)\bp} & 0                               & \eps_{(B_1|B_3}a_{\bp|B_3)C_2}  & 0              & & \\
& P_{C_1C_2}                 & & 0                              & \eps_{(C_1|C_3}a_{A_3\bp|C_2)}  & \eps_{(C_1|C_3}a_{\bp B_3|C_2)} & 0              & & \\
\hline
& 2Q_{A_1}                   & & \eps_{A_1A_3}a_{\bp B_3\bp}    & \eps_{A_1A_3}a_{\bp\bp C_3}    & a_{A_1B_3C_3}                   & a_{A_1 \bp\bp} & & \\

& 2Q_{B_1}                   & & \eps_{B_1B_3}a_{A_3\bp\bp}    & a_{A_3B_1C_3}                   & -\eps_{B_1B_3}a_{\bp\bp C_3}     & -a_{\bp B_1\bp} & & \\

& 2Q_{C_1}                   & & a_{A_3B_3C_1}                  & -\eps_{C_1C_3}a_{A_3\bp\bp}      & -\eps_{C_1C_3}a_{\bp B_3\bp}    & a_{\bp\bp C_1} & &
\end{tabular}
\end{ruledtabular}
\end{table*}

The norm squared $\braket{\Psi}{\Psi}$ is given by
\begin{equation}
\begin{split}
\langle \Psi | \Psi \rangle &=(-)^{X_1+Y_1+Z_1}\delta^{X_1X_2}\delta^{Y_1Y_2}\delta^{Z_1Z_2}a_{X_1Y_1Z_1}^{\#}a_{X_2Y_2Z_2}\\
&=\delta^{A_1A_2} \delta^{B_1B_2}\delta^{C_1C_2}a_{A_1B_1C_1}^{\#}a_{A_2B_2C_2} \\
&\phantom{=}- \delta^{A_1A_2}\delta^{B_1B_2}a_{ A_1B_1\bp}^{\#}a_{A_2B_2\bp}\\
&\phantom{=}- \delta^{A_1A_2}\delta^{C_1C_2}a_{A_1\bp C_1}^{\#}a_{A_2\bp C_2}\\
&\phantom{=}- \delta^{B_1B_2}\delta^{C_1C_2}a_{\bp B_1 C_1}^{\#}a_{\bp B_2 C_2}\\
&\phantom{=}+ \delta^{A_1A_2}a_{ A_1\bp\bp}^{\#}a_{A_2\bp\bp}\\
&\phantom{=}+ \delta^{B_1B_2}a_{\bp B_1 \bp }^{\#}a_{\bp B_2\bp }\\
&\phantom{=}+ \delta^{C_1C_2}a_{\bp \bp C_1}^{\#}a_{\bp \bp C_2}\\
&\phantom{=}- a_{\bp\bp\bp}^{\#}a_{\bp\bp\bp},
\end{split}
\end{equation}
where $\bra{\Psi}= (\ket{\Psi})^\ddag$ and $\langle \Psi | \Psi \rangle$ is the conventional inner product which is manifestly $uOSp(1|2)_A\times uOSp(1|2)_B \times uOSp(1|2)_C$ invariant.

The three-superqubit density matrix is given by
\begin{equation}
\begin{split}
\rho&=\ketbra{\Psi}{\Psi}\\
&=(-)^{X_2+Y_2+Z_2}\ket{X_1Y_1Z_1}a_{X_1Y_1Z_1}a_{X_2Y_2Z_2}^\#\bra{X_2Y_2Z_2}.
\end{split}
\end{equation}
The singly reduced density matrices are defined using the
partial supertraces
\begin{equation}
\begin{split}
\rho_{AB}&=\sum_Z(-)^Z\bra{Z}\rho\ket{Z},\\
\rho_{BC}&=\sum_X(-)^X\bra{X}\rho\ket{X},\\
\rho_{CA}&=\sum_Y(-)^Y\bra{Y}\rho\ket{Y},
\end{split}
\end{equation}
or
\begin{equation}
\begin{split}
\rho_{AB}&=\sum_Z(-)^{X_2+Y_2} \ket{X_1Y_1}a_{X_1Y_1Z}a_{X_2Y_2Z}^\#\bra{X_2Y_2} ,\\
\rho_{BC}&=\sum_X(-)^{Y_2+Z_2} \ket{Y_1Z_1}a_{XY_1Z_1}a_{XY_2Z_2}^\#\bra{Y_2Z_2},\\
\rho_{CA}&=\sum_Y(-)^{X_2+Z_2} \ket{X_1Z_1}a_{X_1YZ_1}a_{X_2YZ_2}^\#\bra{X_2Z_2}.
\end{split}
\end{equation}
The doubly reduced density matrices for Alice, Bob, and Charlie are given by the partial supertraces
\begin{equation}
\begin{split}
\rho_A&=\sum_{Y,Z}(-)^{Y+Z}\bra{YZ}\rho\ket{YZ},\\
\rho_B&=\sum_{X,Z}(-)^{X+Z}\bra{XZ}\rho\ket{XZ},\\
\rho_C&=\sum_{X,Y}(-)^{X+Y}\bra{XY}\rho\ket{XY},
\end{split}
\end{equation}
or
\begin{equation}
\begin{split}
\rho_A&=\sum_{Y,Z}(-)^{X_2} \ket{X_1}a_{X_1YZ}a_{X_2YZ}^\#\bra{X_2},\\
\rho_B&=\sum_{X,Z}(-)^{Y_2} \ket{Y_1}a_{XY_1Z}a_{XY_2Z}^\#\bra{Y_2},\\
\rho_C&=\sum_{X,Y}(-)^{Z_2} \ket{Z_1}a_{XYZ_1}a_{XYZ_2}^\#\bra{Z_2}.
\end{split}
\end{equation}

\section{Super entanglement}
\label{sec:superentanglement}

\subsection{Two superqubits}

In seeking a supersymmetric generalization of the 2-tangle \eqref{eq:2-tangle} one might be tempted to replace the determinant of $a_{AB}$ by the Berezinian of $a_{XY}$
\begin{equation}
\Ber a_{XY}=\det(a_{AB}-a_{A \bp}a_{\bp\bp}^{-1}a_{\bp B})a_{\bp\bp}^{-1}.
\end{equation}
See \hyperref[sec:toolkit]{Appendix~\ref*{sec:toolkit}}. However, although the Berezinian is the natural supersymmetric extension of the determinant, it is not defined for vanishing $a_{\bp\bp}$, making it unsuitable as an entanglement measure.

A better candidate follows from writing
\begin{equation}
\begin{split}
\det a_{AB}=\half a^{AB}a_{AB}&=\half\tr(a^t\eps a\eps^t)\\
                              &=\half\tr[(a\eps)^t\eps a],
\end{split}
\end{equation}
This expression may be generalized by  a straightforward promotion of the trace and transpose to the supertrace and supertranspose and replacing the $SL(2)$ invariant tensor $\eps$ with the $OSp(1|2)$ invariant tensor $E$.  See \hyperref[sec:toolkit]{Appendix~\ref*{sec:toolkit}}. This yields a quadratic polynomial, which we refer to as the superdeterminant, denoted $\sdet$:
\begin{equation}\label{eq:2superqubitinv}
\begin{gathered}
\sdet a_{XY}=\half\str[(aE)^{st}Ea]\\
\begin{split}
&=\half(a^{AB}a_{AB}-a^{A\bp}a_{A\bp}-a^{\bp B}a_{\bp B}-a^{\bp\bp}a_{\bp\bp})\\
&=(a_{00}a_{11}-a_{01}a_{10}+a_{0\bp}a_{1\bp}+a_{\bp0}a_{\bp1})-\half a_{\bp\bp}{}^2,
\end{split}
\end{gathered}
\end{equation}
which is clearly not equal to the Berezinian, but is nevertheless supersymmetric since $Q_A$ annihilates $a^{AB}a_{AB}-a^{\bp B}a_{\bp B}$ and $a^{A\bp}a_{A\bp}+a^{\bp\bp}a_{\bp\bp}$, while $Q_B$ annihilates $a^{AB}a_{AB}-a^{A\bp}a_{A\bp}$ and $a^{\bp B}a_{\bp B}+a^{\bp\bp}a_{\bp\bp}$. Satisfyingly, \eqref{eq:2superqubitinv} reduces to $\det a_{AB}$ when $a_{A\bp}$, $a_{\bp B}$, and $a_{\bp\bp}$ are set to zero. We then define the super 2-tangle as
 \begin{equation}
\tau_{XY}=4 \sdet a_{XY} (\sdet a_{XY})^\#.
\end{equation}
In summary, 2-superqubit entanglement seems to have the same two entanglement classes as 2-qubits with the invariant $\det a_{AB}$ replaced by its supersymmetric counterpart $\sdet a_{XY}$.

Non-superentangled states are given by product states for which
$a_{AB}=a_Ab_B$, $a_{A\bp}=a_Ab_{\bp}$, $a_{\bp B}=a_{\bp}b_B$, $a_{\bp\bp}=a_{\bp}b_{\bp}$, and $\sdet a_{XY}$ vanishes. This provides a nontrivial consistency check.

An example of a normalized physical superentangled state is given by
\begin{equation}
\ket{\Psi} = \tfrac{1}{\sqrt{3}}(\ket{00} +\ket{11} +i\ket{\bp\bp})
\end{equation}
for which
\begin{equation}
\sdet a_{XY}=\tfrac{1}{3}+\half\cdot \tfrac{1}{3}=\half
\end{equation}
and
\begin{equation}
\tau_{XY}=4 \sdet a_{XY} (\sdet a_{XY})^\#=1.
\end{equation}
So this state is not only entangled but maximally entangled,  just like the Bell state
\begin{equation}
\ket{\Psi} = \tfrac{1}{\sqrt{2}}(\ket{00} +\ket{11} )
\end{equation}
for which $\sdet a_{XY}=1/2$ and $\tau_{XY}=1$.  Another more curious example is
\begin{equation}
\ket{\Psi} = i\ket{\bp\bp}
\end{equation}
which is not a product state since $a_{\bp\bp}$ is pure body and hence could never be formed by the product of two odd Grassmann numbers. In fact $\sdet a_{XY}=1/2$ and $\tau_{XY}=1$, so this state is also maximally entangled.

\begin{figure}[ht]
\centering
\includegraphics[width=.5\textwidth]{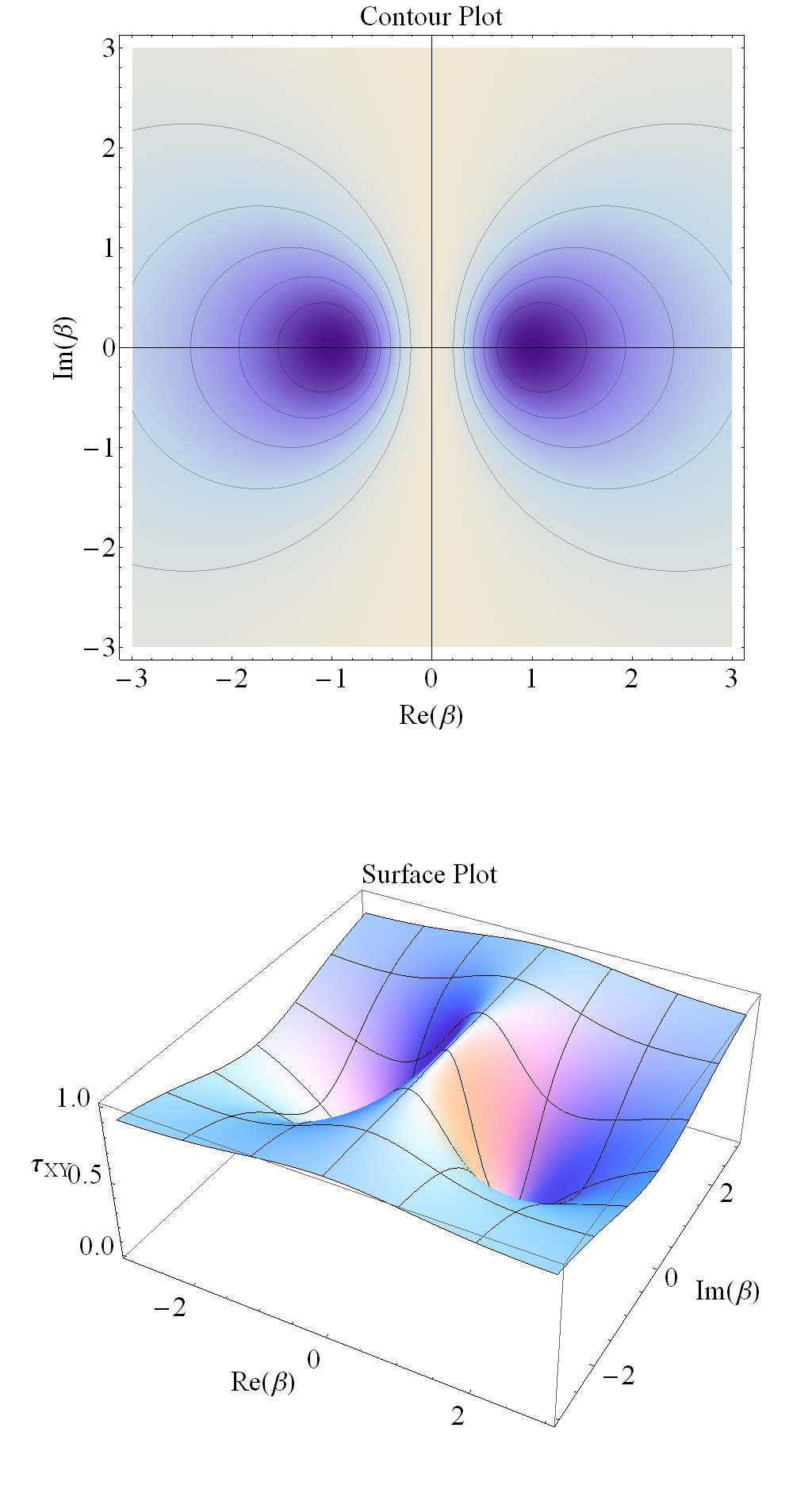}
\caption{The 2-tangle $\tau_{XY}$ for the state \eqref{eq:examplestate} for a complex parameter $\beta$.\label{fig:example}}
\end{figure}

\begin{figure}[ht]
\centering
\includegraphics[width=0.5\textwidth]{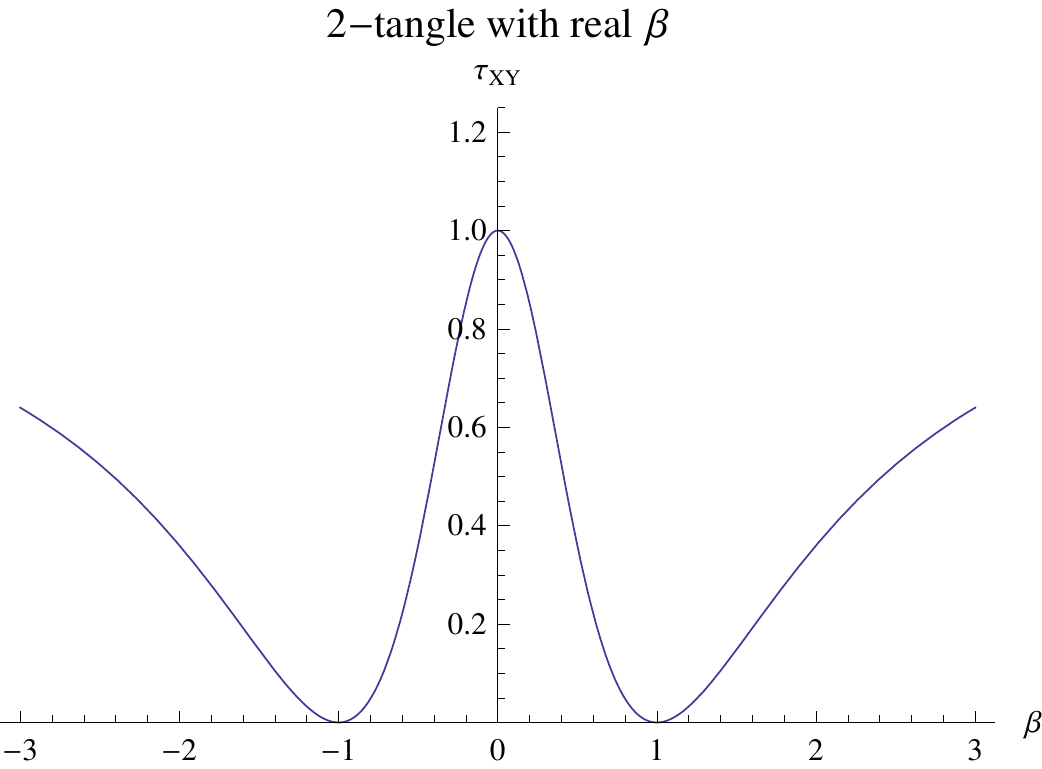}
\caption{The 2-tangle $\tau_{XY}$ for the state \eqref{eq:examplestate} for a real parameter $\beta$.\label{fig:realexample}}
\end{figure}

We may interpolate between these two examples with the normalized state
\begin{equation}\label{eq:examplestate}
(|\alpha|^2+|\beta|^2)^{-1/2}[\alpha\ket{\Psi}_{\text{Bell}}+\beta\ket{\bp\bp}],
\end{equation}
where $\alpha, \beta\in\mathds{C}$, for which we have
\begin{equation}
\begin{split}
\sdet{a_{XY}}&=\frac{1}{2}\frac{\alpha^2-\beta^2}{|\alpha|^2+|\beta|^2},\\
\tau_{XY}&=\frac{|\alpha^2-\beta^2|^2}{(|\alpha|^2+|\beta|^2)^2}.
\end{split}
\end{equation}
The entanglement for this state is displayed as a function of the complex parameter $\beta$ in \autoref{fig:example} for the case $\alpha=1$.
Note, in particular, that while the entanglement is maximized for arbitrary pure imaginary $\beta$, it has its minimum value on the real axis at $\beta=\pm1$ as shown in \autoref{fig:realexample}.

\subsection{Three superqubits}

In seeking to generalize the 3-tangle \eqref{eq:3tangle}, invariant under $[SL(2)]^3$, to a supersymmetric object, invariant under $[OSp(1|2)]^3$, we need to find a quartic polynomial that reduces to Cayley's hyperdeterminant  when $a_{AB\bp}$, $a_{A\bp C}$, $a_{\bp BC}$, $a_{A\bp \bp}$,  $a_{\bp B \bp}$, $a_{\bp \bp C }$, and $a_{\bp \bp \bp }$ are set to zero. We do this by generalizing the $\gamma$ matrices:
\begin{subequations}
\begin{gather}
\begin{split}
\gamma_{A_1A_2} :=&\phantom{-\ } a_{A_1}{}^{BC} a_{A_2BC} - a_{A_1}{}^{B\bp}a_{A_2B\bp}\\
                  &- a_{A_1}{}^{\bp C}a_{A_2\bp C} - a_{A_1}{}^{\bp\bp} a_{A_2\bp\bp},
\end{split}\\
\begin{split}
\gamma_{A_1\bp} :=&\phantom{-\ } a_{A_1}{}^{BC} a_{\bp BC} + a_{A_1}{}^{B\bp}a_{\bp B\bp}\\
                  &+ a_{A_1}{}^{\bp C}a_{\bp\bp C} - a_{A_1}{}^{\bp\bp} a_{\bp\bp\bp},
\end{split}\\
\begin{split}
\gamma_{\bp A_2} :=&\phantom{-\ } a_{\bp}{}^{BC} a_{A_2BC} - a_{\bp}{}^{B\bp}a_{A_2B\bp}\\
                   &-a_{\bp}{}^{\bp C}a_{A_2\bp C} - a_{\bp}{}^{\bp\bp} a_{A_2\bp\bp},
\end{split}
\end{gather}
\end{subequations}
together with their $B$ and $C$ counterparts; notice that the building blocks with two indices are bosonic and those  with one index are fermionic. The final bosonic possibility, $\gamma_{(\bp \bp)}$, vanishes identically.
The simple supersymmetry relations are given by
\begin{equation}
\begin{split}
Q_{A_1}\gamma_{A_2A_3} &= \eps_{A_1(A_2} \gamma_{A_3)\bp}\\
Q_{A_1}\gamma_{A_2\bp} &= \half \gamma_{A_1A_2}\\
Q_{B}\gamma_{A_1A_2}   &= 0 = Q_{C}\gamma_{A_1A_2}\\
Q_{B} \gamma_{A\bp}    &= 0 = Q_{C} \gamma_{A\bp}.
\end{split}
\end{equation}
Using these expressions we define the superhyperdeterminant, denoted $\sDet a$:
\begin{equation}
\sDet a_{XYZ} = \half(\gamma^{A_1A_2}\gamma_{A_1A_2} - \gamma^{A\bp}\gamma_{A\bp}-\gamma^{\bp A}\gamma_{\bp A})
\label{eq:sDet}
\end{equation}
which is invariant under the action of the superalgebra.  The corresponding expressions singling out superqubits $B$ and $C$ are also invariant and equal to \eqref{eq:sDet}. $\sDet a_{XYZ}$ can be seen as the definition of the super-Cayley determinant of the cubic superhypermatrix given in \autoref{fig:superhypermatrix}.

Writing
\begin{equation}
\Gamma^A:=\begin{pmatrix}\gamma_{A_1A_2}&\vrule&\gamma_{A_1\bp}\\ \hline\gamma_{\bp A_2}&\vrule&\gamma_{\bp\bp}\end{pmatrix}=
\begin{pmatrix}\gamma_{A_1A_2}&\vrule&\gamma_{A_1\bp}\\ \hline\gamma_{A_2\bp}&\vrule&0\end{pmatrix},
\end{equation}
we obtain an invariant analogous to \eqref{eq:2superqubitinv}
\begin{equation}
\sDet a_{XYZ}=\half\str[(\Gamma^A E)^{st}E\Gamma^A]
\end{equation}
so that
\begin{equation}
\sDet a_{XYZ}=-\sdet\Gamma^A
\end{equation}
in analogy to the conventional three-qubit identity \eqref{eq:hyperequalsdetgamma}. This result for sDet agrees with that of  \cite{Castellani:2009pi}.

Finally, using $\Gamma^A$ we are able to define the supersymmetric generalization $T_{XYZ}$ of the 3-qubit tensor $T_{ABC}$ as defined in \eqref{eq:Tofgamma},
\begin{equation}
T_{XYZ}=\Gamma^{A}_{XX'}a\indices{^{X'}_{YZ}}.
\end{equation}
It is not difficult to verify that $T_{XYZ}$ transforms in precisely the same way as $a_{XYZ}$ (as given in \autoref{tab:action3}) under $\mathfrak{osp}(1|2)\oplus \mathfrak{osp}(1|2)\oplus \mathfrak{osp}(1|2)$. The superhyperdeterminant may then also be written as
\begin{equation}
\begin{split}
\sDet a_{XYZ}&=T_{ABC}a^{ABC}+T_{\bp BC}a^{\bp BC}\\
&\phantom{=}-T_{A\bp C}a^{A\bp C}-T_{AB\bp}a^{AB\bp}\\
&\phantom{=}-T_{A\bp\bp}a^{A\bp\bp}+T_{\bp B\bp}a^{\bp B\bp}\\
&\phantom{=}+T_{\bp\bp C}a^{\bp\bp C}-T_{\bp\bp\bp}a^{\bp\bp\bp}.
\end{split}
\end{equation}

In this sense $\sDet a_{XYZ}$, $(\Gamma^{A})_{X_1X_2}$, and $T_{XYZ}$ are the natural supersymmetric generalizations of the hyperdeterminant, $\Det a_{ABC}$, and the covariant tensors, $(\gamma^A)_{A_1A_2}$ and $T_{ABC}$, of the conventional 3-qubit treatment summarized in \autoref{sec:3qubitentanglement}.  Finally we are in a position to define the super 3-tangle:
\begin{equation}
\tau_{XYZ}= 4 \sqrt{\sDet a_{XYZ} (\sDet a_{XYZ})^\#.}
\end{equation}
In summary 3-superqubit entanglement seems to have the same five entanglement classes as that of 3-qubits shown in \autoref{tab:merge}, with the covariants
$a_{ABC}, \gamma^A,\gamma^B, \gamma^C, T_{ABC}$ and $\Det a_{ABC}$ replaced by their supersymmetric counterparts $a_{XYZ}, \Gamma^A,\Gamma^B, \Gamma^C, T_{XYZ}$ and $\sDet a_{ABC} $.

Completely separable nonsuperentangled states are given by product states for which $a_{ABC}=a_Ab_Bc_C, a_{AB\bp}=a_Ab_Bc_{\bp}, a_{A\bp C}=a_Ab_{\bp}c_C, a_{\bp BC}=a_{\bp}b_Bc_C, a_{A\bp \bp}=a_Ab_{\bp}c_{\bp}, a_{\bp B \bp}=a_{\bp}b_Bc_{\bp}, a_{\bp \bp C}=a_{\bp}b_ {\bp}c_C, a_ {\bp\bp\bp}=a_{\bp}b_ {\bp}c_{\bp}$, and $\sDet a_{XYZ}$ vanishes.  This provides a nontrivial consistency check.

An example of a normalized physical biseparable state is provided by
\begin{equation}
\ket{\Psi} = \tfrac{1}{\sqrt{3}}(\ket{000}+\ket{011}+\ket{0\bp\bp})			
\end{equation}
for which
\begin{equation}
(\Gamma^{A})_{00}= \tfrac{1}{3}
\end{equation}
and $\Gamma^{B}$, $\Gamma^{C}$, $T_{XYZ}$ and $\sDet a_{XYZ}$ vanish. More generally, one can consider the combination
\begin{equation}
\ket{\Psi} = (|\alpha|^2+|\beta|^2)^{-1/2}[\tfrac{1}{\sqrt{2}}\alpha(\ket{000}+\ket{011})+\beta\ket{0\bp\bp}]		
\end{equation}
for which
\begin{equation}
(\Gamma^{A})_{00}= \frac{\alpha^2-\beta^2}{|\alpha|^2+|\beta|^2}
\end{equation}
and the other covariants vanish.

An example of a normalized physical W state is provided by
\begin{equation}
\begin{split}
\ket{\Psi} = \tfrac{1}{\sqrt{6}}(&\ket{110}+\ket{101}+\ket{011}\\
					            +&\ket{\bp\bp 1}+\ket{\bp 1\bp}+\ket{1\bp\bp})
\end{split}
\end{equation}
for which
\begin{equation}
(\Gamma^{A})_{11}=(\Gamma^{B})_{11}=(\Gamma^{C})_{11}= -\half
\end{equation}
and
\begin{equation}
T_{111}=\tfrac{1}{2\sqrt{6}}
\end{equation}
while $\sDet a_{XYZ}$ vanishes. One could also consider
\begin{equation}
\begin{split}
\ket{\Psi} = \tfrac{1}{\sqrt{3}}(|\alpha|^2+|\beta|^2)^{-1/2}[&\alpha(\ket{110}+\ket{101}+\ket{011})\\
					 +&\beta(\ket{\bp\bp 1}+\ket{\bp 1\bp}+\ket{1\bp\bp})]
\end{split}
\end{equation}
for which
\begin{equation}
(\Gamma^{A})_{11}=(\Gamma^{B})_{11}=(\Gamma^{C})_{11}= -\frac{2\alpha^2+\beta^2}{3(|\alpha|^2+|\beta|^2)}
\end{equation}
and
\begin{equation}
T_{111}=\frac{\alpha(2\alpha^2+\beta^2)}{3\sqrt{3}(|\alpha|^2+|\beta|^2)^{3/2}}
\end{equation}
while the other $T$ components and $\sDet a_{XYZ}$ vanish.

An example of a normalized physical superentangled state is provided by
\begin{equation}
\begin{split}
\ket{\Psi} = \tfrac{1}{\sqrt{8}}(&\ket{000}+\ket{\bp\bp 0}+\ket{\bp 0\bp}+\ket{0\bp\bp}\\
					 +&\ket{111}+\ket{\bp\bp 1}+\ket{\bp 1\bp}+\ket{1\bp\bp})
\end{split}
\end{equation}
for which
\begin{equation}
\sDet a_{XYZ}= \tfrac{1}{64}
\end{equation}
and
\begin{equation}
\tau_{XYZ}=4 \sqrt{\sDet a_{XYZ} (\sDet a_{XYZ})^\#}=\tfrac{1}{16}.
\end{equation}

\section{Conclusion}

In this paper we have taken the first steps toward generalizing quantum information theory to  super quantum information theory. We introduced the superqubit defined over an appropriate super Hilbert space. We acknowledge that there are still important issues to address, notably how to interpret ``physical'' states with nonvanishing soul for which probabilities are no longer real numbers but elements of a Grassmann algebra.  (The sum of the probabilities still add up to one, however.) The examples of \autoref{sec:superentanglement} avoided this problem, being pure body.  DeWitt advocates retaining only such pure body states in the Hilbert space \cite{DeWitt:1984}, but this may be too draconian. See \cite{rudolph-2000-214} for an alternative approach.

Nevertheless, for the SLOCC equivalence group $[SL(2,\cmplxs)]^n$  and the LOCC equivalence group $[SU(2)]^n$, we presented their minimal supersymmetric extensions,  $[OSp(1|2)]^n$ and $[uOSp(1|2)]^n$ respectively, and showed explicitly how superqubits would transform under these groups for $n=1,2,3$. Furthermore, we found supersymmetric invariants that are the obvious candidates  for supersymmetric entanglement measures for $n=2,3$. We hope in future work to classify fully the 2 and 3 superqubit entanglement classes and their corresponding orbits as was done for the 2 and 3 qubit entanglement classes in \cite{Dur:2000, Miyake:2002, Borsten:2009yb}.

As noted in the \hyperref[sec:intro]{Introduction}, a physical realization of our superqubit is more likely to be found in condensed-matter physics than high-energy physics. While the polarizations of a photon or the spins of an electron provide examples of a qubit, the inclusions of photinos or selectrons do not obviously provide examples of a superqubit, since the supersymmetrization of the (S)LOCC equivalence groups is distinct from the supersymmetrization of the spacetime Poincaré group.

We would also like to point out that this work is part of the ongoing correspondence between ideas in string and M-theory and ideas in quantum information theory. See \cite{Borsten:2008wd} for a review. This paper continues the trend of using mathematical tools from one side to describe phenomena on the other.

\begin{acknowledgments}

We are grateful to L. Castellani, P. A. Grassi, and L. Sommovigo for useful correspondence, to S. Bellucci for helpful discussions, and to Kazuki Hasebe for alerting us to the supersymmetric $t$-$J$ model. This work was supported in part by the STFC under rolling Grant No. ST/G000743/1.

\end{acknowledgments}

\section*{Note added}

A very interesting paper has recently appeared \cite{Castellani:2010yz}, which analyzes  quantum computing with superqubits.

\appendix

\section{Superlinear algebra}
\label{sec:toolkit}

Grassmann numbers are the $2^n$-dimensional vectors populating the Grassmann algebra $\Lambda_n$, which is generated by $n$ mutually anticommuting elements $\{\theta^i\}_{i=1}^n$.

Any Grassmann number $z$ may be decomposed into ``body'' $z_\mathcal{B}\in\mathds{C}$ and ``soul'' $z_\mathcal{S}$ viz.
\begin{equation}
\begin{split}
z&=z_\mathcal{B}+z_\mathcal{S}\\
z_\mathcal{S}&=\textstyle\sum_{k=1}^\infty\tfrac{1}{k!}c_{a_1\cdots a_k}\theta^{a_1}\cdots\theta^{a_k},
\end{split}
\end{equation}
where $c_{a_1\cdots a_k}\in\mathds{C}$ are totally antisymmetric. For finite dimension $n$ the sum terminates at $k=2^n$ and the soul is nilpotent $z_{\mathcal{S}}^{n+1}=0$.

One may also decompose $z$ into even and odd parts $u$ and $v$
\begin{equation}
\begin{split}
u&=z_\mathcal{B}+\textstyle\sum_{k=1}^\infty\tfrac{1}{(2k)!}c_{a_1\cdots a_{2k}}\theta^{a_1}\cdots\theta^{a_{2k}}\\
v&=\textstyle\sum_{k=0}^\infty\tfrac{1}{(2k+1)!}c_{a_1\cdots a_{2k+1}}\theta^{a_1}\cdots\theta^{a_{2k+1}},
\end{split}
\end{equation}
which may also be expressed as the direct sum decomposition $\Lambda_n=\Lambda_{n}^{0}\oplus\Lambda_{n}^{1}$. Furthermore, analytic functions $f$ of Grassmann numbers are defined via
\begin{equation}\label{eq:analyticgrassmann}
f(z):=\sum_{k=0}^\infty\tfrac{1}{k!}f^{(k)}(z_\mathcal{B})z_\mathcal{S}^k,
\end{equation}
where $f^{(k)}(z_{\mathcal{B}})$ is the $k^{\textrm{th}}$ derivative of $f$ evaluated at $z_{\mathcal{B}}$ and is well defined if $f$ is nonsingular at $z_{\mathcal{B}}$ \cite{DeWitt:1984}.

One defines the \emph{grade of a Grassmann number} as
\begin{equation}
\deg x:=\begin{cases}0&x\in\Lambda_{n}^{0}\\1&x\in\Lambda_{n}^{1},\end{cases}
\end{equation}
where the grades 0 and 1 are referred to as even and odd, respectively.

Define the star $^\star$ and superstar $^\#$ operators \cite{Scheunert:1977, Berezin:1981,Frappat:2000} satisfying the following properties:
\begin{equation}\label{eq:cond1}
\begin{gathered}
\begin{aligned}
(\Lambda_n^0)^\star&=\Lambda_n^0,&(\Lambda_n^1)^\star&=\Lambda_n^1,\\
(\Lambda_n^0)^\#&=\Lambda_n^0,&(\Lambda_n^1)^\#&=\Lambda_n^1,
\end{aligned}\\
\begin{aligned}
(x \theta_i)^\star&=x^* \theta_{i}^\star,& \theta_{i}^{\star\star}&=\theta_{i},&(\theta_i\theta_j)^\star&=\theta_{j}^\star\theta_{i}^\star,\\
(x \theta_i)^\#&=x^* \theta_{i}^\#,&\theta_{i}^{\#\#}&=-\theta_{i},&(\theta_i\theta_j)^\#&=\theta_{i}^\#\theta_{j}^\#,
\end{aligned}
\end{gathered}
\end{equation}
where $x\in\mathds{C}$ and $^*$ is ordinary complex conjugation, which means
\begin{align}\label{eq:cond2}
\alpha^{\star\star}&=\alpha, & \alpha^{\#\#}&=(-)^{\deg\alpha}\alpha
\end{align}
for pure even/odd Grassmann $\alpha$. The impure case follows by linearity.

Following \cite{DeWitt:1984} one may, if so desired, take the formal limit $n\to\infty$ defining the infinite dimensional vector space $\Lambda_\infty$. Elements of $\Lambda_\infty$ are called \emph{supernumbers}. Our results are independent of the dimension of the underlying Grassmann algebra and one can use supernumbers throughout, but for the sake of simplicity we restrict to finite dimensional algebra by assigning just one Grassmann generator $\theta$ and its superconjugate $\theta^\#$ to every superqubit.

The grade definition applies to the components $T_{X_1\cdots X_k}$ of any $k$-index array of Grassmann numbers $T$, but one may also define $\deg X_i$, the \emph{grade of an index}, for such an array by specifying a characteristic function from the range of the index $X_i$ to the set $\{0,1\}$. In general the indices can have different ranges and the characteristic functions can be arbitrary for each index. It is then possible to define $\deg T$, the \emph{grade of an array}, as long as the compatibility condition
\begin{equation}
\deg T\equiv\deg(T_{X_1\cdots X_k})+\sum_{i=1}^k \deg X_i\mod2\quad\forall\ X_i
\end{equation}
is satisfied. In precisely such cases the entries of $T$ satisfy
\begin{equation}
\begin{gathered}
\deg(T_{X_1\cdots X_k})=\deg T+\sum_{i=1}^k\deg X_i\mod2,\\
\implies\deg T=\deg(T_{\underbrace{1\cdots1}_k}),\\
\deg(T_1T_2)=\deg T_1+\deg T_2\mod2,
\end{gathered}
\end{equation}
so that in other words $T$ is partitioned into blocks with definite grade such that the nearest neighbors of any block are of the opposite grade to that block. The array grade simply distinguishes the two distinct ways of accomplishing such a partition (i.e. the two possible grades of the first element $T_{1\cdots1}$). Grassmann numbers and the Grassmann number grade may be viewed as special cases of arrays and the array grade.

Special care must be taken not to confuse this notion of array grade with whether the array entries at even/odd index positions vanish. An array $T$ may be decomposed as
\begin{equation}
T=T_E+T_O,
\end{equation}
where the pure even part $T_{E}$ is obtained from $T$ by setting to zero all entries satisfying $\deg(T_{X_1\cdots X_k})=1$, and similarly \emph{mutatis mutandis} for $T_{O}$. The property of being pure even or pure odd is therefore independent of the array grade as defined above.

The various grades commonly appear in formulae as powers of -1 and the shorthand
\begin{equation}
(-)^X:=(-1)^{\deg X}
\end{equation}
is often used. The indices of superarrays may be supersymmetrized as follows:
\begin{equation}
\begin{gathered}
T_{X_1\cdots \llbracket X_i|\cdots|X_j\rrbracket\cdots X_k}:=\\
\half[T_{X_1\cdots X_i\cdots X_j\cdots X_k}+(-)^{X_iX_j}T_{X_1\cdots  X_j\cdots X_i\cdots X_k}].
\end{gathered}
\end{equation}
While we require these definitions for some of our considerations, one typically only uses arrays with 0, 1, or 2 indices where the characteristic functions are monotonic: supernumbers, supervectors, and supermatrices, respectively. Functions of grades extend to mixed superarrays (with nonzero even \emph{and} odd parts) by linearity.

A $(p|q)\times(r|s)$ supermatrix is just an $(p+q)\times(r+s)$-dimensional block partitioned matrix
\begin{equation}
M=\kbordermatrix{&r&\vrule&s\\p&A&\vrule&B\\ \hline q&C&\vrule&D}
\end{equation}
where entries in the $A$ and $D$ blocks are grade $\deg M$, and those in the $B$ and $C$ blocks are grade $\deg M+1\mod2$. The special cases $s=0$ or $q=0$ can be permitted to make the definition encapsulate row and column supervectors. Supermatrix multiplication is defined as for ordinary matrices; however, the trace, transpose, adjoint, and determinant have distinct super versions \cite{Frappat:2000,Varadarajan:2004}.

The supertrace $\str M$ of a supermatrix is $M$ defined as
\begin{equation}
\str M:=\sum_X(-)^{(X+M)X}M_{XX}
\end{equation}
and is linear, cyclic modulo sign, and insensitive to the supertranspose
\begin{equation}
\begin{split}
\str(M+N)&=\str(M)+\str(N)\\
\str(MN)&=(-)^{MN}\str(NM)\\
\str M^{st}&=\str M.
\end{split}
\end{equation}

The supertranspose $M^{st}$ of a supermatrix $M$ is defined componentwise as
\begin{equation}
M^{st}{}_{X_1X_2}:=(-)^{(X_2+M)(X_1+X_2)}M_{X_2X_1}.
\end{equation}
Unlike the transpose the supertranspose is not idempotent; instead,
\begin{equation}
\begin{split}
M^{st\,st}{}_{X_1X_2}&=(-)^{(X_1+X_2)}M_{X_1X_2},\\
M^{st\,st\,st}{}_{X_1X_2}&=(-)^{(X_1+M)(X_1+X_2)}M_{X_2X_1},\\
M^{st\,st\,st\,st}{}_{X_1X_2}&=M_{X_1X_2},
\end{split}
\end{equation}
so that it is of order 4. The supertranspose also satisfies
\begin{equation}
(MN)^{st}=(-)^{MN}N^{st}M^{st}.
\end{equation}

The adjoint $^\dag$ and superadjoint $^\ddag$ of a supermatrix are defined as
\begin{equation}
\begin{split}
M^\dag&:=M^{\star t}\\
M^\ddag&:=M^{\# st},
\end{split}
\end{equation}
and satisfy
\begin{equation}
\begin{aligned}
M^{\dag\dag}&=M,&M^{\ddag\ddag}&=(-)^{M}M,\\
(MN)^\dag&=N^\dag M^\dag,&(MN)^\ddag&=(-)^{MN}N^\ddag M^\ddag.
\end{aligned}
\end{equation}

The preservation of anti-super-Hermiticity, $M^\ddag=-M$, under scalar multiplication by Grassmann numbers, as required for the proper definition of $\mathfrak{uosp}(1|2)$ \cite{Schunck:2004ck}, necessitates the left/right multiplication rules:
\begin{equation}\label{eq:scalarmultiplication}
\begin{split}
(\alpha M)_{X_1X_2}&=(-)^{X_1\alpha}\alpha M_{X_1X_2},\\
(M\alpha)_{X_1X_2}&=(-)^{X_2\alpha}M_{X_1X_2}\alpha.
\end{split}
\end{equation}
The Berezinian is defined as
\begin{equation}\label{eq:Ber}
\begin{split}
\Ber M&:=\det(A-BD^{-1}C)/\det(D)\\
      &=\det(A)/\det( D - C A^{-1} B)
\end{split}
\end{equation}
and is multiplicative, insensitive to the supertranspose, and generalizes the relationship between trace and determinant
\begin{equation}
\begin{split}
\Ber(MN)&=\Ber(M)\Ber(N)\\
\Ber M^{st}&=\Ber M\\
\Ber e^M&=e^{\str M}.
\end{split}
\end{equation}

The  direct sum and super tensor product are unchanged from their ordinary versions. As such, the dimension of the tensor product of two superqubits is given by
\begin{equation}
(2|1)\otimes(2|1)=(2|1|2|3|1),
\end{equation}
while the threefold product is
\begin{equation}
(2|1)^{\otimes3}=(2|1|2|3|3|1|2|3|1|2|1|2|3|1),
\end{equation}
with similar results holding for the associated density matrices. In analogy with the ordinary case we have
\begin{equation}
\begin{split}
(M\otimes N)^t&=M^t\otimes N^t\\
(M\otimes N)^{st}&=M^{st}\otimes N^{st}\\
\str(M\otimes N)&=\str M\str N.
\end{split}
\end{equation}
These definitions are manifestly compatible with Hermiticity and super-Hermiticity.

Denoting the total number of bosonic elements in the product of $n$ superqubits by $B_n$, and similarly the total number of fermionic elements by $F_n$, we know that $B_n$ ($F_n$) is given by the total number of basis kets with an even (odd) number of $\bp$'s:
\begin{equation}
\begin{split}
B_n&={n\choose 0}2^n+{n\choose 2}2^{n-2}+\cdots=\frac{3^n+1}{2}\\
F_n&={n\choose 1}2^{n-1}+{n\choose 3}2^{n-3}+\cdots=\frac{3^n-1}{2}
\end{split}
\end{equation}
so that, in particular, $B_n-F_n=1$: the number of bosonic elements is always one more than the number of fermionic ones.

In supermatrix representations of superalgebras, one may represent the superbracket of generators $M$ and $N$ as
\begin{equation}
\llbracket M,N\rrbracket:=MN-N_EM-N_O(M_E-M_O).
\end{equation}
One may also consider supermatrices $M$ and $N$ whose components are themselves supermatrices. Provided the component supermatrices are pure even (odd) at even (odd) index positions (e.g. $M_{11}$ is a pure even supermatrix for even $M$), one may write the superbracket of such supermatrices as
\begin{equation}
\begin{gathered}
\llbracket M_{X_1X_2},N_{X_3X_4}\rrbracket=\\
M_{X_1X_2}N_{X_3X_4}-(-)^{(X_1+X_2)(X_3+X_4)}N_{X_3X_4}M_{X_1X_2},
\end{gathered}
\end{equation}
where the final two indices are suppressed. This grouping of supermatrices into supermatrices is useful for summarizing the superbrackets of superalgebras.

\section{Orthosymplectic superalgebras}\label{sec:osp}

Supermatrix representations of the orthosymplectic supergroup $OSp(p|2q)$ consist of supermatrices $M\in GL(p|2q)$ satisfying
\begin{equation}\label{eq:ospgroupdef}
M^{st}EM=E,
\end{equation}
but for convenience we choose instead to use supermatrices $M\in GL(2q|p)$ satisfying \eqref{eq:ospgroupdef}. In this convention, the invariant supermatrix $E$ is defined by
\begin{align}
E&:=\begin{pmatrix}\mathds{J}_{2q}&\vrule&0\\ \hline0&\vrule&\mathds{1}_p\end{pmatrix},&
\mathds{J}_{2q}&:=\begin{pmatrix}0&\mathds{1}_q\\-\mathds{1}_q&0\end{pmatrix}.
\end{align}
Definitions of supermatrices, the supertranspose, and further details of superlinear algebra may be found in \hyperref[sec:toolkit]{Appendix~\ref*{sec:toolkit}}.

Writing a generic supermatrix $\mathfrak{M}$ of the super Lie algebra $\mathfrak{osp}(p|2q)$ as
\begin{equation}
\mathfrak{M}=\begin{pmatrix}A&\vrule&B\\ \hline C&\vrule&D\end{pmatrix}
\end{equation}
permits \eqref{eq:ospgroupdef} to be rewritten as the following conditions on the blocks of the algebra supermatrices:
\begin{align}\label{eq:ospblockconditions}
A^t\mathds{J}&=-\mathds{J}A, & C&=B^t\mathds{J}, & D^t&=-D.
\end{align}
Depending on the value of $p$, the superalgebra falls into one of three basic, ``classical'' families
\begin{equation}
\mathfrak{osp}(p|2q)=
\begin{cases}
B(r,q) & p=2r+1,\ r\geq0\\
C(q+1) & p=2\\
D(r,q) & p=2r,\ r\geq2.
\end{cases}
\end{equation}
Clearly it is the first case that will concern us, in particular, with $r=0,q=1$. $B(r,q)$ has rank $q+r$, dimension $2(q+r)^2+3q+r$, and even part $\mathfrak{so}(p)\oplus\mathfrak{sp}(2q)$, which for $\mathfrak{osp}(1|2)$ are 1, 5, and $\mathfrak{sl}(2)$, respectively.

One generates $\mathfrak{osp}(p|2q)$ as a matrix superalgebra by defining the supermatrices $U$ and $G$
\begin{gather}
\begin{gathered}\label{eq:Gdef}
(U_{X_1X_2})_{X_3X_4}:=\delta_{X_1X_4}\delta_{X_2X_3},\\
G:=\begin{pmatrix}\mathds{J}_{2q}&\vrule&0\\ \hline 0&\vrule&H_p\end{pmatrix},
\end{gathered}
\shortintertext{where}
H_p:=\begin{cases}\phantom{[}\sigma_1\otimes\mathds{1}_r&p=2r\\ [\sigma_1\otimes\mathds{1}_r]\oplus(1)&p=2r+1\end{cases}
\end{gather}
with $\sigma_1$ being the first Pauli matrix. Here the indices $X_i$ range from 1 to $2q+p$ and are partitioned as $X_i=(\bar{X}_i,\dot{X}_i)$ with $\bar{X}_i$ ranging from 1 to $2q$, and $\dot{X}_i$ taking on the remaining $p$ values. Note that under \eqref{eq:Gdef}, $G$ has the following symmetry properties
\begin{equation}
\begin{gathered}
\begin{aligned}
G_{\bar{X}_1\bar{X}_2}&=-G_{\bar{X}_2\bar{X}_1},&G_{\dot{X}_1\dot{X}_2}&=+G_{\dot{X}_2\dot{X}_1},
\end{aligned}\\
G_{\bar{X}_1\dot{X}_2}=0=G_{\dot{X}_2\bar{X}_1},
\end{gathered}
\end{equation}
which are shared with the invariant supermatrix $E$. In the special case $p=1$, $G$ reduces to $E$.

The generators $T$ are obtained as
\begin{equation}
T_{X_1X_2}=2G_{\llbracket X_1|X_3}U_{X_3|X_2 \rrbracket},
\end{equation}
where $T$ has array grade zero and the index grades are monotonically increasing:
\begin{equation}
\deg X:=\begin{cases}0&X\in\{1,\dotsc,2q\}\\1&X\in\{2q+1,\dotsc,2q+p\}.\end{cases}
\end{equation}
Clearly $T$ has symmetry properties $T_{X_1X_2}=T_{\llbracket X_1X_2\rrbracket}$. The $2q(2q+1)/2$ generators $T_{\bar{X}_1\bar{X}_2}$ generate $\mathfrak{sp}(2q)$, the $p(p-1)/2$ generators $T_{\dot{X}_1\dot{X}_2}$ generate $\mathfrak{so}(p)$, and both are even (bosonic), while the $2pq$ generators $T_{\bar{X}_1\dot{X}_2}$ are odd (fermionic). These supermatrices yield the $\mathfrak{osp}(p|2q)$ superbrackets
\begin{equation}
\llbracket T_{X_1X_2},T_{X_3X_4}\rrbracket:=4G_{\llbracket X_1\llbracket X_3}T_{X_2\rrbracket X_4\rrbracket},
\end{equation}
where the supersymmetrization on the right-hand side is over pairs $X_1X_2$ and $X_3X_4$ as on the left-hand side. The action of the generators on $(2q|p)$-dimensional supervectors $a_X$ is given by
\begin{equation}
(T_{X_1X_2})_{X_3X_4}a_{X_4}\equiv(T_{X_1X_2}a)_{X_3}=2G_{\llbracket X_1|X_3}a_{X_2\rrbracket}
\end{equation}
This action may be generalized to an $N$-fold super tensor product of $(2q|p)$ supervectors by labeling the indices with integers $k=1,2,\dotsc,N$
\begin{equation}
\begin{gathered}
(T_{X_kY_k}a)_{Z_1\cdots Z_k\cdots Z_N}=\\
(-)^{(X_k+Y_k)\sum_{i=1}^{k-1}|Z_i|}2G_{\llbracket X_k|Z_k}a_{Z_1\cdots |Y_k\rrbracket\cdots Z_N}.
\end{gathered}
\end{equation}
In our special case $p=1$ we denote the lone dotted index $\dot{X}_i$ by a bullet $\bp$ and start counting the barred indices at zero so that $X_i=(0,1,\bp)$. Obviously the $T_{\bp\bp}$ generator vanishes identically, leaving only the following superbrackets:
\begin{equation}
\begin{split}
\left[T_{A_1A_2},T_{A_3A_4}\right]              &= 4E_{(A_1(A_3}T_{A_2)A_4)} \\
\left[T_{A_1A_2},T_{A_3\bp}\right]              &= 2E_{(A_1|A_3}T_{A_2)\bp} \\
\left\lbrace T_{A_1\bp},T_{A_2\bp}\right\rbrace &= T_{A_1A_2},
\end{split}
\end{equation}
which are written out in \autoref{tab:Talgebra} with $T_A\equiv T_{A\bp}\equiv T_{\bp A}$.
\begin{table}
\caption{$\mathfrak{osp}(1|2)$ superbrackets.\label{tab:Talgebra}}
\begin{ruledtabular}
\begin{tabular*}{\textwidth}{@{\extracolsep{\fill}}*{8}{M{c}}}
&        & T_{01}   & T_{00}   & T_{11}  & T_{0}  & T_{1}  \\
\hline
& T_{01} & 0        & -2T_{00} & 2T_{11} & -T_0   & T_1    \\
& T_{00} & 2T_{00}  & 0        & 4T_{01} & 0      & 2T_0   \\
& T_{11} & -2T_{11} & -4T_{01} & 0       & -2T_1  & 0      \\
& T_{0}  & T_0      & 0        & 2T_1    & T_{00} & T_{01} \\
& T_{1}  & -T_1     & -2T_0    & 0       & T_{01} & T_{11}
\end{tabular*}
\end{ruledtabular}
\end{table}
Explicitly the generators are
\begin{equation}
\begin{gathered}
T_{01}=\begin{pmatrix}-1&0&\vrule&0\\0&1&\vrule&0\\ \hline 0&0&\vrule&0\end{pmatrix},\\
\begin{aligned}
T_{00}&=\begin{pmatrix}0&2&\vrule&0\\0&0&\vrule&0\\ \hline 0&0&\vrule&0\end{pmatrix},&
T_{11}&=\begin{pmatrix}0&0&\vrule&0\\-2&0&\vrule&0\\ \hline 0&0&\vrule&0\end{pmatrix},
\end{aligned}\\
\begin{aligned}
T_{0}&=\begin{pmatrix}0&0&\vrule&1\\0&0&\vrule&0\\ \hline 0&1&\vrule&0\end{pmatrix},&
T_{1}&=\begin{pmatrix}0&0&\vrule&0\\0&0&\vrule&1\\ \hline -1&0&\vrule&0\end{pmatrix}.
\end{aligned}
\end{gathered}
\end{equation}
In order to make contact with \cite{Castellani:2009pi}, we rescale the generators into a new supermatrix $P$
\begin{align}
P_{X_1X_2}:=\half T_{X_1X_2}\equiv E_{\llbracket X_1|X_3}U_{X_3|X_2 \rrbracket}
\end{align}
to yield the superbrackets
\begin{equation}\label{eq:ospbrackets}
\begin{split}
\left[ P_{A_1A_2}, P_{A_3A_4} \right]       &= 2\eps_{(A_1(A_3} P_{A_2)A_4)}  \\
\left[P_{A_1A_2}, Q_{A_3}\right]            &= \eps_{(A_1|A_3} Q_{A_2)} \\
\left\lbrace Q_{A_1}, Q_{A_2} \right\rbrace &= \half P_{A_1A_2},
\end{split}
\end{equation}
where $Q_A\equiv P_A$, which are summarized as
\begin{equation}
\llbracket P_{X_1X_2}, P_{X_3X_4}\rrbracket  = 2E_{\llbracket X_1\llbracket X_3} P_{X_2\rrbracket X_4\rrbracket}.
\end{equation}
The rescaled generators have the action
\begin{equation}
\begin{gathered}
(P_{X_1X_2}a)_{X_3}=E_{\llbracket X_1|X_3}a_{X_2\rrbracket}\\
(P_{X_kY_k}a)_{Z_1\cdots Z_k\cdots Z_N}=\\
(-)^{(X_k+Y_k)\sum_{i=1}^{k-1}Z_i}E_{\llbracket X_k|Z_k}a_{Z_1\cdots |Y_k\rrbracket\cdots Z_N},
\end{gathered}
\end{equation}
which summarizes Tables \hyperref[tab:action1]{\ref*{tab:action1}}, \hyperref[tab:action2]{\ref*{tab:action2}} and \hyperref[tab:action3]{\ref*{tab:action3}}.


\begin{thebibliography}{10}%
\makeatletter
\providecommand \@ifxundefined [1]{%
 \ifx #1\undefined \expandafter \@firstoftwo
 \else \expandafter \@secondoftwo
\fi
}%
\providecommand \@ifnum [1]{%
 \ifnum #1\expandafter \@firstoftwo
 \else \expandafter \@secondoftwo
\fi
}%
\providecommand \enquote [1]{``#1''}%
\providecommand \bibnamefont  [1]{#1}%
\providecommand \bibfnamefont [1]{#1}%
\providecommand \citenamefont [1]{#1}%
\providecommand\href[0]{\@sanitize\@href}%
\providecommand\@href[1]{\endgroup\@@startlink{#1}\endgroup\@@href}%
\providecommand\@@href[1]{#1\@@endlink}%
\providecommand \@sanitize [0]{\begingroup\catcode`\&12\catcode`\#12\relax}%
\@ifxundefined \pdfoutput {\@firstoftwo}{%
 \@ifnum{\z@=\pdfoutput}{\@firstoftwo}{\@secondoftwo}%
}{%
 \providecommand\@@startlink[1]{\leavevmode}%
 \providecommand\@@endlink[0]{}%
}{%
 \providecommand\@@startlink[1]{%
  \leavevmode
  \pdfstartlink
   attr{/Border[0 0 1 ]/H/I/C[0 1 1]}%
   user{/Subtype/Link/A<</Type/Action/S/URI/URI(#1)>>}%
  \relax
 }%
 \providecommand\@@endlink[0]{\pdfendlink}%
}%
\providecommand \url  [0]{\begingroup\@sanitize \@url }%
\providecommand \@url [1]{\endgroup\@href {#1}{\urlprefix}}%
\providecommand \urlprefix [0]{URL }%
\providecommand \Eprint[0]{\href }%
\@ifxundefined \urlstyle {%
  \providecommand \doi [1]{doi:\discretionary{}{}{}#1}%
}{%
  \providecommand \doi [0]{doi:\discretionary{}{}{}\begingroup
  \urlstyle{rm}\Url }%
}%
\providecommand \doibase [0]{http://dx.doi.org/}%
\providecommand \Doi[1]{\href{\doibase#1}}%
\providecommand \bibAnnote [3]{%
  \BibitemShut{#1}%
  \begin{quotation}\noindent
    \textsc{Key:}\ #2\\\textsc{Annotation:}\ #3%
  \end{quotation}%
}%
\providecommand \bibAnnoteFile [2]{%
  \IfFileExists{#2}{\bibAnnote {#1} {#2} {\input{#2}}}{}%
}%
\providecommand \typeout [0]{\immediate \write \m@ne }%
\providecommand \selectlanguage [0]{\@gobble}%
\providecommand \bibinfo [0]{\@secondoftwo}%
\providecommand \bibfield [0]{\@secondoftwo}%
\providecommand \translation [1]{[#1]}%
\providecommand \BibitemOpen[0]{}%
\providecommand \bibitemStop [0]{}%
\providecommand \bibitemNoStop [0]{.\EOS\space}%
\providecommand \EOS [0]{\spacefactor3000\relax}%
\providecommand \BibitemShut [1]{\csname bibitem#1\endcsname}%
\bibitem{Coffman:1999jd}%
  \BibitemOpen
  \bibfield{author}{%
  \bibinfo {author} {\bibfnamefont{V.}~\bibnamefont{Coffman}}, \bibinfo
  {author} {\bibfnamefont{J.}~\bibnamefont{Kundu}},\ and\ \bibinfo {author}
  {\bibfnamefont{W.~K.}\ \bibnamefont{Wootters}},\ }%
  \bibfield{journal}{%
  \Doi{10.1103/PhysRevA.61.052306}{\bibinfo {journal} {Phys. Rev.}}\ }%
  \textbf{\bibinfo {volume} {A61}},\ \bibinfo {pages} {052306} (\bibinfo {year}
  {2000}),\
  \Eprint{http://arxiv.org/abs/quant-ph/9907047}{arXiv:quant-ph/9907047}%
  \bibAnnoteFile{NoStop}{Coffman:1999jd}%
\bibitem{Dur:2000}%
  \BibitemOpen
  \bibfield{author}{%
  \bibinfo {author} {\bibfnamefont{W.}~\bibnamefont{{D\"ur}}}, \bibinfo
  {author} {\bibfnamefont{G.}~\bibnamefont{Vidal}},\ and\ \bibinfo {author}
  {\bibfnamefont{J.~I.}\ \bibnamefont{Cirac}},\ }%
  \bibfield{journal}{%
  \Doi{10.1103/PhysRevA.62.062314}{\bibinfo {journal} {Phys. Rev.}}\ }%
  \textbf{\bibinfo {volume} {A62}},\ \bibinfo {pages} {062314} (\bibinfo {year}
  {2000}),\
  \Eprint{http://arxiv.org/abs/quant-ph/0005115}{arXiv:quant-ph/0005115}%
  \bibAnnoteFile{NoStop}{Dur:2000}%
\bibitem{Acin:2001}%
  \BibitemOpen
  \bibfield{author}{%
  \bibinfo {author} {\bibfnamefont{A.}~\bibnamefont{Acin}}, \bibinfo {author}
  {\bibfnamefont{A.}~\bibnamefont{Andrianov}}, \bibinfo {author}
  {\bibfnamefont{E.}~\bibnamefont{Jane}},\ and\ \bibinfo {author}
  {\bibfnamefont{R.}~\bibnamefont{Tarrach}},\ }%
  \bibfield{journal}{%
  \Doi{10.1088/0305-4470/34/35/301}{\bibinfo {journal} {J. Phys.}}\ }%
  \textbf{\bibinfo {volume} {A34}},\ \bibinfo {pages} {6725} (\bibinfo {year}
  {2001}),\
  \Eprint{http://arxiv.org/abs/quant-ph/0009107}{arXiv:quant-ph/0009107}%
  \bibAnnoteFile{NoStop}{Acin:2001}%
\bibitem{Miyake:2002}%
  \BibitemOpen
  \bibfield{author}{%
  \bibinfo {author} {\bibfnamefont{A.}~\bibnamefont{Miyake}}\ and\ \bibinfo
  {author} {\bibfnamefont{M.}~\bibnamefont{Wadati}},\ }%
  \bibfield{journal}{%
  \bibinfo {journal} {Quant. Info. Comp.}\ }%
  \textbf{\bibinfo {volume} {2 (Special)}},\ \bibinfo {pages} {540} (\bibinfo
  {year} {2002}),\
  \Eprint{http://arxiv.org/abs/quant-ph/0212146}{arXiv:quant-ph/0212146}%
  \bibAnnoteFile{NoStop}{Miyake:2002}%
\bibitem{Miyake:2003}%
  \BibitemOpen
  \bibfield{author}{%
  \bibinfo {author} {\bibfnamefont{A.}~\bibnamefont{Miyake}},\ }%
  \bibfield{journal}{%
  \Doi{10.1103/PhysRevA.67.012108}{\bibinfo {journal} {Phys. Rev.}}\ }%
  \textbf{\bibinfo {volume} {A67}},\ \bibinfo {pages} {012108} (\bibinfo {year}
  {2003}),\
  \Eprint{http://arxiv.org/abs/quant-ph/0206111}{arXiv:quant-ph/0206111}%
  \bibAnnoteFile{NoStop}{Miyake:2003}%
\bibitem{Borsten:2009yb}%
  \BibitemOpen
  \bibfield{author}{%
  \bibinfo {author} {\bibfnamefont{L.}~\bibnamefont{Borsten}}, \bibinfo
  {author} {\bibfnamefont{D.}~\bibnamefont{Dahanayake}}, \bibinfo {author}
  {\bibfnamefont{M.~J.}\ \bibnamefont{Duff}}, \bibinfo {author}
  {\bibfnamefont{W.}~\bibnamefont{Rubens}},\ and\ \bibinfo {author}
  {\bibfnamefont{H.}~\bibnamefont{Ebrahim}},\ }%
  \bibfield{journal}{%
  \Doi{10.1103/PhysRevA.80.032326}{\bibinfo {journal} {Phys. Rev.}}\ }%
  \textbf{\bibinfo {volume} {A80}},\ \bibinfo {pages} {032326} (\bibinfo {year}
  {2009}),\ \Eprint{http://arxiv.org/abs/0812.3322}{arXiv:0812.3322
  [quant-ph]}%
  \bibAnnoteFile{NoStop}{Borsten:2009yb}%
\bibitem{Castellani:2009pi}%
  \BibitemOpen
  \bibfield{author}{%
  \bibinfo {author} {\bibfnamefont{L.}~\bibnamefont{Castellani}}, \bibinfo
  {author} {\bibfnamefont{P.~A.}\ \bibnamefont{Grassi}},\ and\ \bibinfo
  {author} {\bibfnamefont{L.}~\bibnamefont{Sommovigo}},\ }%
  \bibfield{journal}{%
  \Doi{10.1016/j.physletb.2009.06.032}{\bibinfo {journal} {Phys. Lett.}}\ }%
  \textbf{\bibinfo {volume} {B678}},\ \bibinfo {pages} {308} (\bibinfo {year}
  {2009}),\ \Eprint{http://arxiv.org/abs/0904.2512}{arXiv:0904.2512 [hep-th]}%
  \bibAnnoteFile{NoStop}{Castellani:2009pi}%
\bibitem{Wiegmann:1988}%
  \BibitemOpen
  \bibfield{author}{%
  \bibinfo {author} {\bibfnamefont{P.~B.}\ \bibnamefont{Wiegmann}},\ }%
  \bibfield{journal}{%
  \Doi{10.1103/PhysRevLett.60.821}{\bibinfo {journal} {Phys. Rev. Lett.}}\ }%
  \textbf{\bibinfo {volume} {60}},\ \bibinfo {pages} {2445} (\bibinfo {month}
  {Jun}\ \bibinfo {year} {1988})%
  \bibAnnoteFile{NoStop}{Wiegmann:1988}%
\bibitem{Sarkar:1991}%
  \BibitemOpen
  \bibfield{author}{%
  \bibinfo {author} {\bibfnamefont{S.}~\bibnamefont{Sarkar}},\ }%
  \bibfield{journal}{%
  \Doi{10.1088/0305-4470/24/5/026}{\bibinfo {journal} {J. Phys.}}\ }%
  \textbf{\bibinfo {volume} {A24}},\ \bibinfo {pages} {1137} (\bibinfo {year}
  {1991})%
  \bibAnnoteFile{NoStop}{Sarkar:1991}%
\bibitem{PhysRevB.46.9234}%
  \BibitemOpen
  \bibfield{author}{%
  \bibinfo {author} {\bibfnamefont{A.}~\bibnamefont{Foerster}}\ and\ \bibinfo
  {author} {\bibfnamefont{M.}~\bibnamefont{Karowski}},\ }%
  \bibfield{journal}{%
  \Doi{10.1103/PhysRevB.46.9234}{\bibinfo {journal} {Phys. Rev. B}}\ }%
  \textbf{\bibinfo {volume} {46}},\ \bibinfo {pages} {9234} (\bibinfo {month}
  {Oct}\ \bibinfo {year} {1992})%
  \bibAnnoteFile{NoStop}{PhysRevB.46.9234}%
\bibitem{Essler:1992nk}%
  \BibitemOpen
  \bibfield{author}{%
  \bibinfo {author} {\bibfnamefont{F.~H.~L.}\ \bibnamefont{Essler}}\ and\
  \bibinfo {author} {\bibfnamefont{V.~E.}\ \bibnamefont{Korepin}}}%
   (\bibinfo {year} {1992}),\
  \Eprint{http://arxiv.org/abs/hep-th/9207007}{arXiv:hep-th/9207007}%
  \bibAnnoteFile{NoStop}{Essler:1992nk}%
\bibitem{Mavromatos:1999xj}%
  \BibitemOpen
  \bibfield{author}{%
  \bibinfo {author} {\bibfnamefont{N.~E.}\ \bibnamefont{Mavromatos}}\ and\
  \bibinfo {author} {\bibfnamefont{S.}~\bibnamefont{Sarkar}},\ }%
  \bibfield{journal}{%
  \Doi{10.1103/PhysRevB.62.3438}{\bibinfo {journal} {Phys. Rev.}}\ }%
  \textbf{\bibinfo {volume} {B62}},\ \bibinfo {pages} {3438} (\bibinfo {year}
  {2000}),\
  \Eprint{http://arxiv.org/abs/cond-mat/9912323}{arXiv:cond-mat/9912323}%
  \bibAnnoteFile{NoStop}{Mavromatos:1999xj}%
\bibitem{Hasebe:2004hy}%
  \BibitemOpen
  \bibfield{author}{%
  \bibinfo {author} {\bibfnamefont{K.}~\bibnamefont{Hasebe}},\ }%
  \bibfield{journal}{%
  \Doi{10.1103/PhysRevLett.94.206802}{\bibinfo {journal} {Phys. Rev. Lett.}}\
  }%
  \textbf{\bibinfo {volume} {94}},\ \bibinfo {pages} {206802} (\bibinfo {year}
  {2005}),\ \Eprint{http://arxiv.org/abs/hep-th/0411137}{arXiv:hep-th/0411137}%
  \bibAnnoteFile{NoStop}{Hasebe:2004hy}%
\bibitem{Arovas:2009dx}%
  \BibitemOpen
  \bibfield{author}{%
  \bibinfo {author} {\bibfnamefont{D.~P.}\ \bibnamefont{Arovas}}, \bibinfo
  {author} {\bibfnamefont{K.}~\bibnamefont{Hasebe}}, \bibinfo {author}
  {\bibfnamefont{X.-L.}\ \bibnamefont{Qi}},\ and\ \bibinfo {author}
  {\bibfnamefont{S.-C.}\ \bibnamefont{Zhang}},\ }%
  \bibfield{journal}{%
  \Doi{10.1103/PhysRevB.79.224404}{\bibinfo {journal} {Phys. Rev.}}\ }%
  \textbf{\bibinfo {volume} {B79}},\ \bibinfo {pages} {224404} (\bibinfo {year}
  {2009}),\ \Eprint{http://arxiv.org/abs/0901.1498}{arXiv:0901.1498
  [cond-mat.str-el]}%
  \bibAnnoteFile{NoStop}{Arovas:2009dx}%
\bibitem{Plenio:2007}%
  \BibitemOpen
  \bibfield{author}{%
  \bibinfo {author} {\bibfnamefont{M.~B.}\ \bibnamefont{Plenio}}\ and\ \bibinfo
  {author} {\bibfnamefont{S.}~\bibnamefont{Virmani}},\ }%
  \bibfield{journal}{%
  \bibinfo {journal} {Quant. Inf. Comp.}\ }%
  \textbf{\bibinfo {volume} {7}},\ \bibinfo {pages} {1} (\bibinfo {year}
  {2007}),\
  \Eprint{http://arxiv.org/abs/quant-ph/0504163}{arXiv:quant-ph/0504163}%
  \bibAnnoteFile{NoStop}{Plenio:2007}%
\bibitem{Horodecki:2007}%
  \BibitemOpen
  \bibfield{author}{%
  \bibinfo {author} {\bibfnamefont{R.}~\bibnamefont{Horodecki}}, \bibinfo
  {author} {\bibfnamefont{P.}~\bibnamefont{Horodecki}}, \bibinfo {author}
  {\bibfnamefont{M.}~\bibnamefont{Horodecki}},\ and\ \bibinfo {author}
  {\bibfnamefont{K.}~\bibnamefont{Horodecki}},\ }%
  \bibfield{journal}{%
  \Doi{10.1103/RevModPhys.81.865}{\bibinfo {journal} {Rev. Mod. Phys.}}\ }%
  \textbf{\bibinfo {volume} {81}},\ \bibinfo {pages} {865} (\bibinfo {month}
  {Jun}\ \bibinfo {year} {2009}),\
  \Eprint{http://arxiv.org/abs/quant-ph/0702225}{arXiv:quant-ph/0702225}%
  \bibAnnoteFile{NoStop}{Horodecki:2007}%
\bibitem{Bennett:1999}%
  \BibitemOpen
  \bibfield{author}{%
  \bibinfo {author} {\bibfnamefont{C.~H.}\ \bibnamefont{Bennett}}, \bibinfo
  {author} {\bibfnamefont{S.}~\bibnamefont{Popescu}}, \bibinfo {author}
  {\bibfnamefont{D.}~\bibnamefont{Rohrlich}}, \bibinfo {author}
  {\bibfnamefont{J.~A.}\ \bibnamefont{Smolin}},\ and\ \bibinfo {author}
  {\bibfnamefont{A.~V.}\ \bibnamefont{Thapliyal}},\ }%
  \bibfield{journal}{%
  \Doi{10.1103/PhysRevA.63.012307}{\bibinfo {journal} {Phys. Rev.}}\ }%
  \textbf{\bibinfo {volume} {A63}},\ \bibinfo {pages} {012307} (\bibinfo {year}
  {2000}),\
  \Eprint{http://arxiv.org/abs/quant-ph/9908073}{arXiv:quant-ph/9908073}%
  \bibAnnoteFile{NoStop}{Bennett:1999}%
\bibitem{Linden:1997qd}%
  \BibitemOpen
  \bibfield{author}{%
  \bibinfo {author} {\bibfnamefont{N.}~\bibnamefont{Linden}}\ and\ \bibinfo
  {author} {\bibfnamefont{S.}~\bibnamefont{Popescu}},\ }%
  \bibfield{journal}{%
  \bibinfo {journal} {Fortschr. Phys.}\ }%
  \textbf{\bibinfo {volume} {46}},\ \bibinfo {pages} {567} (\bibinfo {year}
  {1998}),\
  \Eprint{http://arxiv.org/abs/quant-ph/9711016}{arXiv:quant-ph/9711016}%
  \bibAnnoteFile{NoStop}{Linden:1997qd}%
\bibitem{Greenberger:1989}%
  \BibitemOpen
  \bibfield{author}{%
  \bibinfo {author} {\bibfnamefont{D.~M.}\ \bibnamefont{Greenberger}}, \bibinfo
  {author} {\bibfnamefont{M.}~\bibnamefont{Horne}},\ and\ \bibinfo {author}
  {\bibfnamefont{A.}~\bibnamefont{Zeilinger}},\ }%
  \emph{\bibinfo {title} {{Bell's Theorem, Quantum Theory and Conceptions of
  the Universe}}}\ (\bibinfo {publisher} {Kluwer Academic},\ \bibinfo {address}
  {Dordrecht},\ \bibinfo {year} {1989})\ ISBN \bibinfo {isbn} {0-7923-0496-9}%
  \bibAnnoteFile{NoStop}{Greenberger:1989}%
\bibitem{Cayley:1845}%
  \BibitemOpen
  \bibfield{author}{%
  \bibinfo {author} {\bibfnamefont{A.}~\bibnamefont{Cayley}},\ }%
  \enquote{\bibinfo {title} {{On the theory of linear transformations}},}\
  \bibinfo {note}
  {\href{http://www.archive.org/download/collmathpapers01caylrich/collmathpape%
rs01caylrich.pdf}{\textit{Camb. Math. J.} \textbf{4} (1845) 193--209}}%
  \bibAnnoteFile{NoStop}{Cayley:1845}%
\bibitem{Sudbery:2001}%
  \BibitemOpen
  \bibfield{author}{%
  \bibinfo {author} {\bibfnamefont{A.}~\bibnamefont{Sudbery}},\ }%
  \bibfield{journal}{%
  \Doi{10.1088/0305-4470/34/3/323}{\bibinfo {journal} {J. Phys.}}\ }%
  \textbf{\bibinfo {volume} {A34}},\ \bibinfo {pages} {643} (\bibinfo {year}
  {2001}),\
  \Eprint{http://arxiv.org/abs/quant-ph/0001116}{arXiv:quant-ph/0001116}%
  \bibAnnoteFile{NoStop}{Sudbery:2001}%
\bibitem{Kempe:1999vk}%
  \BibitemOpen
  \bibfield{author}{%
  \bibinfo {author} {\bibfnamefont{J.}~\bibnamefont{Kempe}},\ }%
  \bibfield{journal}{%
  \Doi{10.1103/PhysRevA.60.910}{\bibinfo {journal} {Phys. Rev.}}\ }%
  \textbf{\bibinfo {volume} {A60}},\ \bibinfo {pages} {910} (\bibinfo {year}
  {1999}),\
  \Eprint{http://arxiv.org/abs/quant-ph/9902036}{arXiv:quant-ph/9902036}%
  \bibAnnoteFile{NoStop}{Kempe:1999vk}%
\bibitem{DeWitt:1984}%
  \BibitemOpen
  \bibfield{author}{%
  \bibinfo {author} {\bibfnamefont{B.}~\bibnamefont{DeWitt}},\ }%
  \emph{\bibinfo {title} {{Supermanifolds}}},\ \bibinfo {edition} {2nd}\ ed.,\
  {Cambridge Monographs on Mathematical Physics}\ (\bibinfo {publisher}
  {Cambridge University Press},\ \bibinfo {year} {1984})%
  \bibAnnoteFile{NoStop}{DeWitt:1984}%
\bibitem{VanProeyen:1999ni}%
  \BibitemOpen
  \bibfield{author}{%
  \bibinfo {author} {\bibfnamefont{A.}~\bibnamefont{Van~Proeyen}}}%
   (\bibinfo {year} {1999}),\
  \Eprint{http://arxiv.org/abs/hep-th/9910030}{arXiv:hep-th/9910030}%
  \bibAnnoteFile{NoStop}{VanProeyen:1999ni}%
\bibitem{Frappat:2000}%
  \BibitemOpen
  \bibfield{author}{%
  \bibinfo {author} {\bibfnamefont{L.}~\bibnamefont{Frappat}}, \bibinfo
  {author} {\bibfnamefont{A.}~\bibnamefont{Sciarrino}},\ and\ \bibinfo {author}
  {\bibfnamefont{P.}~\bibnamefont{Sorba}},\ }%
  \emph{\bibinfo {title} {Dictionary on Lie Algebras and Superalgebras}}\
  (\bibinfo {publisher} {Academic Press},\ \bibinfo {year} {2000})\ ISBN
  \bibinfo {isbn} {0-1226-5340-8},\
  \Eprint{http://arxiv.org/abs/hep-th/9607161}{arXiv:hep-th/9607161}%
  \bibAnnoteFile{NoStop}{Frappat:2000}%
\bibitem{Berezin:1981}%
  \BibitemOpen
  \bibfield{author}{%
  \bibinfo {author} {\bibfnamefont{F.~A.}\ \bibnamefont{Berezin}}\ and\
  \bibinfo {author} {\bibfnamefont{V.~N.}\ \bibnamefont{Tolstoy}},\ }%
  \bibfield{journal}{%
  \Doi{10.1007/BF01942332}{\bibinfo {journal} {Commun. Math. Phys.}}\ }%
  \textbf{\bibinfo {volume} {78}},\ \bibinfo {pages} {409} (\bibinfo {year}
  {1981})%
  \bibAnnoteFile{NoStop}{Berezin:1981}%
\bibitem{rudolph-2000-214}%
  \BibitemOpen
  \bibfield{author}{%
  \bibinfo {author} {\bibfnamefont{O.}~\bibnamefont{Rudolph}},\ }%
  \bibfield{journal}{%
  \Doi{10.1007/s002200000281}{\bibinfo {journal} {Commun. Math. Phys.}}\ }%
  \textbf{\bibinfo {volume} {214}},\ \bibinfo {pages} {449} (\bibinfo {year}
  {2000}),\
  \Eprint{http://arxiv.org/abs/math-ph/9910047}{arXiv:math-ph/9910047}%
  \bibAnnoteFile{NoStop}{rudolph-2000-214}%
\bibitem{Borsten:2008wd}%
  \BibitemOpen
  \bibfield{author}{%
  \bibinfo {author} {\bibfnamefont{L.}~\bibnamefont{Borsten}}, \bibinfo
  {author} {\bibfnamefont{D.}~\bibnamefont{Dahanayake}}, \bibinfo {author}
  {\bibfnamefont{M.~J.}\ \bibnamefont{Duff}}, \bibinfo {author}
  {\bibfnamefont{H.}~\bibnamefont{Ebrahim}},\ and\ \bibinfo {author}
  {\bibfnamefont{W.}~\bibnamefont{Rubens}},\ }%
  \bibfield{journal}{%
  \Doi{10.1016/j.physrep.2008.11.002}{\bibinfo {journal} {Phys. Rep.}}\ }%
  \textbf{\bibinfo {volume} {471}},\ \bibinfo {pages} {113} (\bibinfo {year}
  {2009}),\ \Eprint{http://arxiv.org/abs/0809.4685}{arXiv:0809.4685 [hep-th]}%
  \bibAnnoteFile{NoStop}{Borsten:2008wd}%
\bibitem{Castellani:2010yz}%
  \BibitemOpen
  \bibfield{author}{%
  \bibinfo {author} {\bibfnamefont{L.}~\bibnamefont{Castellani}}, \bibinfo
  {author} {\bibfnamefont{P.~A.}\ \bibnamefont{Grassi}},\ and\ \bibinfo
  {author} {\bibfnamefont{L.}~\bibnamefont{Sommovigo}}}%
   (\bibinfo {year} {2010}),\
  \Eprint{http://arxiv.org/abs/1001.3753}{arXiv:1001.3753 [hep-th]}%
  \bibAnnoteFile{NoStop}{Castellani:2010yz}%
\bibitem{Scheunert:1977}%
  \BibitemOpen
  \bibfield{author}{%
  \bibinfo {author} {\bibfnamefont{M.}~\bibnamefont{Scheunert}}, \bibinfo
  {author} {\bibfnamefont{W.}~\bibnamefont{Nahm}},\ and\ \bibinfo {author}
  {\bibfnamefont{V.}~\bibnamefont{Rittenberg}},\ }%
  \bibfield{journal}{%
  \Doi{10.1063/1.523149}{\bibinfo {journal} {J. Math. Phys.}}\ }%
  \textbf{\bibinfo {volume} {18}},\ \bibinfo {pages} {155} (\bibinfo {year}
  {1977})%
  \bibAnnoteFile{NoStop}{Scheunert:1977}%
\bibitem{Varadarajan:2004}%
  \BibitemOpen
  \bibfield{author}{%
  \bibinfo {author} {\bibfnamefont{V.~S.}\ \bibnamefont{Varadarajan}},\ }%
  \emph{\bibinfo {title} {{Supersymmetry for mathematicians: an
  introduction}}}\ (\bibinfo {publisher} {American Mathematical Society},\
  \bibinfo {year} {2004})\ ISBN \bibinfo {isbn} {0-8218-3574-2}%
  \bibAnnoteFile{NoStop}{Varadarajan:2004}%
\bibitem{Schunck:2004ck}%
  \BibitemOpen
  \bibfield{author}{%
  \bibinfo {author} {\bibfnamefont{A.~F.}\ \bibnamefont{Schunck}}\ and\
  \bibinfo {author} {\bibfnamefont{C.}~\bibnamefont{Wainwright}},\ }%
  \bibfield{journal}{%
  \Doi{10.1063/1.1850363}{\bibinfo {journal} {J. Math. Phys.}}\ }%
  \textbf{\bibinfo {volume} {46}},\ \bibinfo {pages} {033511} (\bibinfo {year}
  {2005}),\ \Eprint{http://arxiv.org/abs/hep-th/0409257}{arXiv:hep-th/0409257}%
  \bibAnnoteFile{NoStop}{Schunck:2004ck}%
\end{thebibliography}
\end{document}